\documentclass[onecolumn,showpacs,preprintnumbers,amsmath,amssymb]{revtex4}
\usepackage{graphicx}
\usepackage{epstopdf}
\usepackage{amstext,amsfonts,amsbsy,amssymb,bbm}
\usepackage{amsmath,amscd,amssymb,latexsym}

\def\Re{\mathop{\rm Re}\nolimits}

\newcommand {\be}[1]{\begin{eqnarray} \mbox{$\label{#1}$}  }
\newcommand{\ee}{\end{eqnarray}}

\newcommand{\pref}[1]{(\ref{#1})}

\newcommand\ie {{\it i.e.}, }

\newcommand\etal {{\it et al. }}

\newcommand{\nn}{\nonumber\\}
\newcommand{\noi}{\noindent}

\newcommand\half{\frac 1 2 }

\newcommand{\dd}[2]{{d{#1}\over d{#2}}}
\newcommand{\ddtwo}[2]{{d^2{#1}\over d{#2}^2}}

\newcommand{\pd}{\partial}

\newcommand{\mean}[1]{\left \langle #1 \right \rangle}

\newcommand{\ket}[1]{|#1\rangle}
\newcommand{\bra}[1]{\langle #1 |}

\newcommand{\av}[1]{\langle #1\rangle}

\newcommand{\com}[2]{\left[ #1, #2 \right]}

\newcommand{\ga}{ {\alpha} }
\newcommand{\gb}{ {\beta} }
\renewcommand{\gg}{\gamma}
\newcommand{\gG}{\Gamma}
\newcommand{\gd}{ {\delta} }
\newcommand{\gD}{ {\Delta} }
\newcommand{\gf}{\phi}
\newcommand{\gw}{ {\omega} }

\newcommand{\gq}{ {\theta} }

\newcommand{\gr}{\rho}

\newcommand{\gk}{\kappa}

\newcommand{\sgn}{{\rm sgn}}

\newcommand{\cq}{\cosh\xi_q}
\newcommand{\sq}{\sinh\xi_q}
\newcommand{\co}{\cosh\xi_0}
\newcommand{\so}{\sinh\xi_0}

\begin{document}

\title{Are there sharp fractional charges in Luttinger liquids?}
\author{Jon Magne Leinaas$^{a}$, Mats Horsdal$^{a,b}$ and T. H.  Hansson$^{c}$}

\affiliation{${(a)}$ Department of Physics, University of Oslo, N-0316 Oslo, Norway}
\affiliation{${(b)}$ NORDITA, AlbaNova University Center, SE-106 91 Stockholm, Sweden}
\affiliation{${(c)}$ Department of Physics, University of Stockholm, Box 6730,S-11385 Stockholm, Sweden}

\date{June 4, 2009}


\begin{abstract} 
We examine charge fractionalization by chiral separation in a one-dimensional fermion system described by Luttinger liquid theory. The focus is on the question of whether the fractional charges are quantum mechanically sharp, and in the analysis we make  a distinction between the global charge, which is restricted by boundary conditions, and the local charge where a background contribution is subtracted. We show,  by way of examples, that fractional charges of arbitrary values, all which are quantum mechanically sharp, can be introduced by different initial conditions.
Since the system is gapless, excitations of arbitrary low frequency contribute to the fluctuations, it is important to make a precise definition of sharp charges, and this we we do by subtraction of the ground state contribution. We very briefly comment on the relevance of our analysis for proposed experiments.
\end{abstract}

\pacs{}
\maketitle

\section{Introduction}

Charge quantization is regarded as a fundamental phenomenon of nature, and it has an overwhelmingly strong empirical support. All elementary particles that have been observed thus come with a charge that is a multiple of the fundamental unit charge. Although charge quantization is generally considered to be a fundamental quantum mechanical effect, there is no unique way to derive it from fundamental principles. There are, however, certain general arguments for such a connection. One set of arguments is based on the (hypothetical) existence of magnetic monopoles. A consistent quantum theory of systems that contain particles with electric and with magnetic charges requires the Dirac quantization condition to be satisfied, and this in turn implies that both electric and magnetic charges must be quantized \cite{Dirac31}.   Furthermore, the natural description of the electromagnetic field in the presence of monopoles is in terms of topologically non-trivial  fibrebundles\cite{WuYang75}.
Another set of arguments is based on the assumption that the gauge group of electromagnetism is not  the translation group $\mathbb R$ but  rather the unitary group $U(1)$, which is compact. This means that the electromagnetic potentials are angular variables, and from that follows that  the  electric charges are integer multiples of a fundamental unit of charge just as a quantum mechanical angular momentum is quantized in units of $\hbar/2$ \cite{Yang70}. This is a natural scenario in certain unified theories where the gauge group of electromagnetism is embedded in a lager non-Abelian symmetry group.
It is interesting that topology is a basic ingredient in both these approaches, and that they are closely connected in that many unified gauge theories have soliton solutions describing magnetic monopoles\cite{Thooft74, Polyakov74}.

In spite of the fundamental character of the charge quantization condition, {\em charge fractionalization} is a meaningful and interesting phenomenon that may take place under certain conditions, typically on the background of a non-trivial quantum many-particle state.  It is then important to appreciate the distinction between fractionalization of the fundamental charge and fractionalization of the expectation value of the charge, which may arise from quantum fluctuations in the charge density. An example of the latter is the charge distribution of a single electron in the presence of two clearly separated positive ions. While the the expectation value of the local charge close to one of the ions will be $e/2$, the fluctuations are large and any local measurement of the charge will yield either 0 or $e$. Charge fractionalization on the other hand, amount to having {\em sharp} local charges, meaning that a suitable local charge measurement will yield $e/2$.

There are two cases of charge fractionalization in condensed matter systems that has caught much attention. One is the case of half-integer charges in certain one-dimensional crystals characterized by a Peierls instability. In a field theoretical description of this system, the fractional charges are associated with soliton excitations \cite{Jackiw76,Su79,Goldstone81}. \footnote{
It should be noted that although the subject of the cited work was to describe the quasi-one-dimensional electron system of polyacetylen, fractional charges cannot be observed in this system.}
The second case has to do with quasi-particle excitations in the two-dimensional electron gas of the (fractional) quantum Hall effect. In this case the excitations appear with fractional charges as well as fractional statistics, and the values they take are determined by the ground state, with different rational values for each plateau of  the Hall conductivity \cite{Laughlin83,Halperin84,Arovas84}.

An important property of the ground states where fractionally charged excitations have been predicted to exist is that there is a gap in the energy spectrum. For the plateau states of the quantum Hall effect this amounts to  the two-dimensional electron gas behaving as an incompressible liquid. The effect of the gap is to suppress low-frequency fluctuations, and this is what makes it possible for the excitations to appear with sharply defined particle properties in the quantum mechanical sense \cite{Kivelson82,Rajaraman82}.

It is against this background, one should view recent theoretical claims of the existence of fractional charges in one-dimensional fermion systems described by Luttinger liquid theory, such as quantum wires or carbon nano-tubes.  In particular, it has been pointed out that fermions that are scattered on impurities \cite{FisherGlazman97} or simply injected into the one-dimensional system \cite{Pham00} will dynamically be separated into a right-handed and a left-handed component, where each of these carry a non-integer fermion number. It has also been argued that these charges are quantum mechanically {\em sharp} observables \cite{Pham00}, which indicate that one could view them as carried by quasi-particles with fractional charge (and statistics). There have also been suggestions of how one should experimentally detect such fractionally charged objects \cite{LeHur08,Berg08}. However, since these systems are gapless and also since the charges depend continuously on an interaction parameter, $g$, the question of whether the fractional charges can be viewed as quantum mechanical sharp is less clear. In particular  the sharpness of the charges seems not protected by any topological argument.

These questions are far from settled, and the motivation for the present paper is to examine more carefully in what sense fractional fermion numbers, created by chiral separation, can be viewed as being quantum mechanically sharp. We analyze the question by following Heinonen and Kohn \cite{HeinonenKohn87} in introducing a distinction between the total {\em local} and {\em global} charges. The global charge is the full charge of the system, including boundary charges in a one-dimensional system with boundaries.
For a compact space, which is here a  circle since we are in one dimension, it is simply the integrated charge, which is the $q=0$ (momentum) component of the charge density.
In the bosonic field theory description, the quantized values of the global charges are restricted by periodicity conditions and can be viewed as topological quantum numbers that are independent of the chiral separation. In the fermion picture they carry information about the fundamental fermions of the theory and are independent of the particle interactions. The local charges on the other hand are insensitive to the periodicity constraints and may take values different from those of the fundamental fermions. They are defined as the $q\to 0$ limit of the charge densities rather than the $q=0$ component.
We have already mentioned the principal difficulties with defining sharp local charges in gapless systems. 
Nevertheless, we shall see that it is possible  to consistently subtract the contribution emanating from the ground state fluctuations, and we shall   call a charge sharp  if there are no additional contributions.

The paper is organized in the following way. We first introduce the one-dimensional model, which in the low-energy approximation is equivalent to the Tomonaga-Luttinger model, and summarize the basic facts about its bosonized form. We next discuss chiral separation of excitations with non-integer fermion numbers by three specific examples. The first one is the situation discussed by Pham et. al. \cite{Pham00}, where a fermion is injected at one of the two Fermi points of the system, and we show how to reproduce the fractional charge values of the chiral components in our formalism. In the next example we follow the approach of Heinonen and Kohn \cite{HeinonenKohn87}, who introduce what they call the local quasi-particle charge by inserting a fundamental particle and then adiabatically turning on the interaction. The third example describes a situation where a local polarization charge is introduced by applying an external potential. We show that in all  these three cases  the chiral components carry fractional charge, but the values differ between the cases. In the first two examples it depends (in a different way) on the coupling strength, while in the last example it also depends on the strengths of the external potential. 

Next we consider the interesting question whether the fractional charges can be considered as sharp. This we do by evaluating the variance of the local charge in the bosonic description, and we focus first on the ground state fluctuations for which we derive an explicit result. The charge fluctuations of the excited states are then evaluated for each of the three examples, and we find that in all three cases the charges are sharp in the meaning that the variance of the charge operator is identical to the ground state variance. This cast some doubt on the conclusion that any of these should be seen as the charge of a basic charge-carrying object of the Luttinger liquid. We conclude with some comments about this and also the possible relevance of our results for proposed experiments. In the appendix we compare the results of the bosonized theory with explicit calculations using free fermions and without making any low-energy approximations.

Note that charge is throughout the paper taken to be dimensionless, and equal to the fermion number of the many-particle system.

\section{The model}
We consider a system of (non-relativistic) spinless fermions in one dimension, with a many particle Hamiltonian of the standard form
\be{ham1}
H=\int dx \psi^\dag(x)(-{\hbar\over {2m}}\ddtwo{}{x})\psi(x)+\half\int dx\int dx' \rho(x) V(x-x') \rho(x')
\ee
where $\psi(x)$ is the fermion field operator, $\gr(x)=\psi^\dag(x)\psi(x)$ is the particle density and $V(x-x')$ a two-particle interaction. The system is confined to a ring of length $L$, which we assume to be much longer than any relevant physical length, and whenever convenient we may therefore take the limit $L\to\infty$. The field operator is assumed to be periodic on the ring, $\psi(x+L)=\psi(x)$.

In momentum space the Hamiltonian takes the form
\be{ham2}
H=\sum_k\,{{\hbar^2}\over{2m}}k^2\,c_k^{\dag}c_k+\frac{1}{2L}\sum_{q,k_1,k_2}V(q)c_{k_1}^{\dag}c_{k_1+q}c_{k_2}^{\dag}c_{k_2-q}
\ee
with 
\be{field}
\psi(x)={1\over{\sqrt L}} \sum_k c_k e^{i k x}\,,\quad c_k={1\over{\sqrt L}} \int_0^L dx\, \psi(x) e^{-i k x}
\ee
and where $k$ takes the discrete values, $k=2n\pi/L$ with $n$ as an integer. 

With $N_0$ particles in the system, the ground state of the non-interacting system defines a full Fermi sea, where all momentum states between the two Fermi points $\pm k_F$ are occupied and other single particle states are empty. The Fermi momentum is related to the particle number by $k_F=N_0\pi/L$. The particle interaction is assumed not to change this picture in an essential way. With the interaction turned on, the system in the ground state should still define a Fermi sea, but now with a smooth transition from full occupation to no occupation for momentum states in an interval $\Delta k$ about the Fermi points $\pm k_F$.  We assume $\Delta k << k_F$ and consider in the following a low energy description of the system which is restricted to processes that excite particles only in this interval. The particle number we also assume to be close to $N_0$ and we define $N$ to measure the particle number {\em relative} to this state.

The low-energy approximation to the Hamiltonian takes the form of the Tomonaga-Luttinger model \cite{Tomonaga50,Luttinger63}
\be{fermham}
H=v_F\hbar \sum_{\chi ,\,k}(\chi k-k_F):c_{\chi,\, k}^{\dag}\,c_{\chi,\, k}:+\frac{1}{2L} \sum_{\chi,\,q}(V_1(q)\rho_{\chi,\,q}\rho_{\chi,\,-q}+V_2(q)\rho_{\chi,\,q}\rho_{-\chi,\,-q})
\ee
where excitations close to the two Fermi points $\pm k_F$ are assigned different values for the quantum number $\chi=\pm 1$. The interaction is now separated in two parts, with $V_1$ as the interaction between pairs of particles close to the same Fermi point, and $V_2$ as the interaction between particles at opposite Fermi points. The form \pref{ham1} of the original Hamiltonian introduces the restriction $V_1=V_2$, however for more general (non-local) interactions the two interaction potentials may be different.  In the low-energy approximation there is no interaction matrix element that will change the relative number of fermions at the two Fermi points. This means that there are two conserved fermion numbers $N_\chi$, that both take integer values. The parameter $v_F$ in the above expression is the Fermi velocity. In the non-interacting case it is given by $\hbar k_F/m$, but in the interacting case it is renormalized by interactions between the (dynamical) particles close to $\pm k_F$ and the particles trapped in the Fermi sea. In the Hamiltonian \pref{fermham} the operators are normal ordered with respect to the non-interacting Fermi sea.  

Although the $k$ quantum number is in the low energy approximation restricted to small deviations from $\pm k_F$, this restriction can be lifted. This is so since the low energy sector of the theory is not affected by the extension of the values of $k$. Without the restriction the model describes two types of fermions, characterized by different values of  $\chi$, both types with linear dispersion. 

The low energy Hamiltonian can be bosonized, in the standard way, by expressing the Fourier components of the charge density operators, for $q\neq 0$, as boson annihilation and creation operators,
\be{aop}
a_q=\sqrt{{2\pi}\over |q|L}\sum_\chi \gq(\chi q)\,\rho_{\chi,\,q}\,,\quad 
a_q^\dag=\sqrt{{2\pi}\over |q|L}\sum_\chi \gq(\chi q)\,\rho_{\chi,\,-q}
\ee
with $\gq(q)$ as the Heaviside step function. The $q=0$ components of the charge densities define  the conserved fermion number and chiral (current) quantum number
\be{NJ}
N=\sum_{\chi}N_\chi=\sum_{k\, \chi}:c^\dag_{\chi,k} \,c_{\chi,k}:\,,\quad\quad J=\sum_{\chi}\chi N_\chi=\sum_{k\, \chi}\chi\, c^\dag_{\chi,k}\,c_{\chi,k}
\ee
The bosonized form of the Hamiltonian is \cite{Haldane81}
\be{boseham}
H&=&{{\pi\hbar}\over{2L}}(v_N N^2+v_JJ^2)
\nn
&+&{\hbar\over 2}\sum_{q\neq 0}|q|\left[
(v_F+\frac{V_{1}(q)}{2\pi\hbar})
(a_q^{\dag}a_q+a_q a_{q}^{\dag})+\frac{V_{2}(q)}{2\pi\hbar}(a_q^{\dag}a_{-q}^{\dag}+a_q a_{-q})\right] 
\ee
when modified relative to the fermionic Hamiltonian \pref{fermham}  by adding terms that are constant or linear in $N$.  The two velocity parameters are
\be{velpar}
v_N=v_F+\frac{1}{2\pi\hbar}(V_1(0)+V_2(0))\,,\quad
v_J=v_F+\frac{1}{2\pi\hbar}(V_1(0)-V_2(0))
\ee
The low energy sector now corresponds to situations where $|q|$ is effectively restricted to values much smaller than $k_F$ and where  $N$ and $J$ have only small deviations from their ground state values $N=J=0$.

The bosonized Hamiltonian is diagonalized by a Bogoliubov transformation of the form
\be{bogo}
a_q=\cosh\xi_q\,b_q+\sinh\xi_q\,b^\dag_{-q}\nn
a^\dag_q=\cosh\xi_q\,b^\dag_q+\sinh\xi_q\,b_{-q}
\ee
where $\xi_q$ is fixed by the relation
\be{xi}
\tanh2\xi_q=-\frac{V_2(q)}{V_1(q)+2\pi\hbar \,v_F}
\ee
In terms of the new bosonic operators the Hamiltonian gets the diagonal form
\be{hamdiag}
H=\sum_{q\neq 0}\hbar\gw_q \,b_q^{\dag}b_q
+ {{\pi\hbar}\over{2L}}(v_N N^2+v_JJ^2)
\ee
with the frequency $\gw_q$ given by
\be{freq}
\gw_q=\sqrt{\left(v_F+\frac{V_1(q)}{2\pi\hbar}\right)^2-\left(\frac{V_2(q)}{2\pi\hbar}\right)^2}\;|q|
\ee

If in the low energy approximation the interaction potentials can be approximated by constants,
 $V_1(q)\approx V_1(0)$ and $V_2(q)\approx V_2(0)$, then the bosonic Hamiltonian can be given the field theoretic form 
\be{hfield}
H={u\over2}\pi\hbar \int_0^L dx \left[g^{-1}\left(\pd_x\Phi\right)^2+g\left(\pd_x\Theta\right)^2\right]
\ee
with $u=\sqrt{v_N v_J}$ and $g=\sqrt{v_J/v_N}$. The fields $\Theta(x)$ and $\Phi(x)$ are related to the bosonic creation and annihilation operators in the following way
\be{fieldop}
\Theta(x)&=&\Theta_0+{x\over L}J-i\sum_{q\neq 0}{1\over\sqrt{2\pi Lg|q|}}(b_q e^{iqx}-b_q^\dag e^{-iqx})\nn
\Phi(x)&=& \Phi_0-{x\over L}N+i\sum_{q\neq 0}\sqrt{{g\over2\pi L|q|}}\,\sgn(q)(b_q e^{iqx}-b_q^\dag e^{-iqx})
\ee
with $\Theta_0$ and $\Phi_0$ are $x$ independent operators.

In this formulation $\Theta$ and $\pd_x\Phi=d\Phi/dx$ (or alternatively $\Phi$ and $\pd_x\Theta$) are regarded as conjugate field variables, with the basic field commutator given as
\be{fieldcom}
\com{\Theta(x)}{\pd_x\Phi}=\com{\Phi(x)}{\pd_x\Theta}=\frac{i}{\pi}\gd(x-y)
\ee
where $\gd(x)$ should be interpreted as the periodic delta-function on the circle of length $L$. The field $\pd_x\Phi$ can be identified with the fermion number density $\gr(x)$ of the original description  while $\pd_x \Theta$ is proportional to the current density $j(x)$.

In the representation \pref{hfield} the $N$ and $J$ dependent part of the Hamiltonian \pref{hamdiag} does not appear explicitly, but is hidden in the {\em zero mode}, which is the non-propagating $q=0$ part of the theory. This mode is linked to topological properties of the fields, which are reflected in the following quasi-periodic condition 
\be{qper}
\Theta(x+L)=\Theta(x)+J\,,\quad \Phi(x+L) = \Phi(x)-N
\ee
where $N$ and $J$ are integers that are restricted by the fermion condition that $N\pm J$ are even. In the field theoretic description the fermions can therefore be interpreted as topological excitations of the $(\Phi,\Theta)$ fields,  and the $N$ and $J$ dependent part of the Hamiltonian \pref{hamdiag} can thus be interpreted as a topological term.

In the expansion of the field operators $\Phi(x)$ and $\Theta(x)$ the $N$ and $J$ dependent terms, but also the constant operators $\Phi_0$ and $\Theta_0$, represent the zero mode. The presence of $\Phi_0$ and $\Theta_0$ is needed in order to satisfy the canonical commutation relations \pref{fieldcom}, and from this follows the following commutators
\be{noncom}
\com{\Theta_0}{N}=-\com{\Phi_0}{J}=-\frac{i}{\pi}
\ee
They show that $\Theta_0$ and $\Phi_0$ generate operators that change the fermion numbers. The two fundamental fermion creation operators, which increase either $N_+$ or $N_-$ by one unit, are expressed in terms of the operators  $\Theta_0$ and $\Phi_0$ as $\exp(i\pi(\Theta_0 + \Phi_0))$ and $\exp(i\pi(\Theta_0 - \Phi_0))$. 

\section{Chiral separation}
The fields $\Theta(x)$ and $\Phi(x)$ satisfy a one-dimensional wave equation, and can be separated in a natural way in two parts, its right- and left-moving components. These components, which can be defined as
\be{chicomp}
\Theta_\pm(x)= \Theta(x)\mp{1\over g}\Phi(x)
\ee
satisfy the linear differential equations
\be{lin}
(\pd_t\pm u \pd_x)\Theta_\pm(x)=0
\ee
The (Fourier) expansion of the chiral fields have the following form
\be{chiralfield}
\Theta_\pm(x)=\Theta_{0\pm}\pm{2x\over g L}Q_\pm-i\sum_{\pm q> 0}{\sqrt{2\over \pi Lg|q|}}(b_q e^{iqx}-b_q^\dag e^{-iqx})
\ee
with the zero mode operators
\be{zerofields}
\Theta_{0\pm}=\Theta_0\mp{1\over g}\Phi_0 \,,\quad
Q_\pm = \half(N\pm g J)
\ee
These operators satisfy the $g$-independent commutation relations
\be{comut}
\com{\Theta_{0i}}{Q_j}=-{i\over\pi} \gd_{ij} 
\ee
with $i,j=\pm$.

The operators $Q_\pm$ have a natural interpretation as the {\em chiral} fermion number operators. For $g\neq 1$ they take non-integer values and for $g=1$ they coincide with the operators $N_\pm$, which take integer values. However, as we shall discuss further there are some complications concerning this interpretation of the operators.  

It is then of interest to have a closer look at the topological sector of the theory, which is described by the zero mode operators. The quasi-periodic conditions satisfied by $\Theta(x)$ and $\Phi(x)$, can in a natural way be interpreted as the condition that  the variables $\Theta_0$ and $\Phi_0$ define a two-dimensional space with the topology of a torus. The precise periodicity condition is given by the identification
\be{percond}
(\Theta_0,\Phi_0)\equiv (\Theta_0+n-m,\Phi_0+n+m)
\ee
with $n$ and $m$ as (independent) integers, and only functions of $\Theta_0$ and $\Phi_0$ that respect this periodicity are to be considered as observables. The periodicity \pref{percond} is dictated by the spectrum of the two operators $N$ and $J$, and it can be interpreted as a condition on the fermion creation operators $\exp(i(\Theta_0\pm\Phi_0))$, demanding that they preserve this spectrum. 

As an important point to note, the distinction between the two types of independent fermion creation operators  matches the separation into the two types of chiralities when  $g=1$. Thus the operators $\exp(i(\Theta_0\pm\Phi_0))$ create fermions with well defined chirality.
However, for $g\neq 1$ that is no longer the case, and the corresponding mismatch can be seen as a conflict between the dynamical separation of the two chiralities and the periodicity requirement. The former determines the form of the operators $\Theta_\pm(x)$ associated with the right- and left-going modes, while the latter  determines the form of the operators $\exp(i(\Theta_0\pm\Phi_0))$ that produce changes in the fermion numbers.
As a consequence of this the operators $\exp(i\Theta_{0\pm})$, which at the formal level create states with {\em fractional} fermion numbers $Q_\pm$, are not acceptable as creation operators since they do not respect the periodicity constraints that the physical Hilbert space states should satisfy. Only certain combinations of these operators satisfy the periodicity conditions and therefore create states with acceptable (integer) fermion numbers.
 In Ref.~\onlinecite{Pham00} it is argued that operators can nevertheless be defined that create chiral excitations with fractional charge and statistics. However, these are well-defined operators only if the constraints introduced by the boundary conditions are lifted, and conclusions based on the representation of these operators are therefore not fully convincing. 

For the further discussion we find it useful to associate {\em two} different quantum numbers $\chi$ and $\Gamma$ with these two sides of the chiral separation. In the original fermion model $\chi$ characterizes the two branches of fermion states, and is therefore linked to the fermion numbers $N_\chi$ which are fixed by the boundary conditions. In the non-interacting case $\chi$ also defines the chirality of the state, so that $\chi=+1$ corresponds to right-moving and $\chi=-1$ corresponds to left-moving modes. 
When interaction is turned on this picture changes. The fermion numbers $N_\chi$ are still conserved, but $\chi$ is no longer a true chirality quantum number, since a right-moving mode will not be a pure $\chi=+1$ mode, but also involve a (small) component of $\chi=-1$. This mixing of the two fermion branches is caused by the interaction, and is represented by the Bogoliubov transformation which diagonalizes the bosonic Hamiltonian. For the interacting system we therefore specify the chiral modes instead  by $\Gamma$, so that $\Gamma=+$ corresponds to the right-going and $\Gamma=-$ to the left-going modes. Obviously for $g=1$ we have $\chi=\Gamma$.

In order to illustrate the meaning of the two quantum numbers, we consider a two-dimensional representation of the many-fermion system. As discussed in Refs.~\onlinecite{Horsdal07,Horsdal08} a two dimensional electron gas in a strong magnetic field, when confined to the lowest Landau level, is equivalent to a one-dimensional system, and with a harmonic confinement potential in the direction orthogonal to the $x$-axis ($y$-direction) the Hamiltonian gets the form \pref{ham1} when mapped to one dimension. In this case the interaction, in its one-dimensional form, is non-local and therefore gives $V_1(x)\neq V_2(x)$. The same correspondence has recently also been applied in Ref.~\onlinecite{Berg08}, where the possibility of detecting charge fractionalization on the edge of a quantum Hall system has been analyzed.  

The mapping between the densities in the one and two dimensional representations is given by
\be{twodim}
\gr_2(x,y)={\cal N}\int dx'\int d\xi e^{-{1\over l^2} [(x-x')^2+{1\over 4}\xi^2-iy\xi]}\gr_1(x'+{\xi\over 2}, x'-{\xi\over 2})
\ee  
where $\cal N$ is a normalization factor, $l$ is the magnetic length of the electron system and $\gr_1$ is the off-diagonal density operator of the one-dimensional system. The above expression shows that $\gr_2(x,y)$ is closely related to the Wigner function $W(x,k)$ of the one-dimensional system, 
\be{Wigner}
\gr_2(x,y)=N'\int dx'\int dk \,e^{-{1\over l^2} [(x-x')^2+(y-kl^2)^2]}W(x',k)
\ee  
However $\gr_2(x,y)$ is, as opposed to $W(x,k)$, a positive (semi-definite) operator for arbitrary $(x,y)$ due to the Gaussian factor in \pref{Wigner}, and has therefore the character of a true density function. The expression also shows that the $y$-direction in the plane is essentially the momentum direction in the phase space of the one-dimensional system, with the identification $y=kl^2$. The two-dimensional description can thus be considered a phase space representation of the one-dimensional system, with a particular definition of the two-dimensional density, and with the Landau-level model as a specific, physical realization.

The filled Fermi sea, in this representation corresponds to a band of integer Landau level filling between two parallel edges, located at $y=\pm k_Fl^2\equiv \pm y_F$.
For the non-interacting system the two-dimensional particle density is explicitly given by the $x$-independent function
\be{fermi}
\gr_2(x,y)={1\over 4\pi l^2}
\left({\rm erf}\left[\frac{y+k_Fl^2}{l^2}\right]-{\rm erf}\left[\frac{y-k_Fl^2}{l^2}\right]\right)
\ee
with erf$(z)$ denoting the error function. When the interaction is turned on, the ground-state density is essentially given by the same function, but with a slight softening of the edges at $\pm y_F$ \cite{Horsdal07}.

If an additional particle is now injected into the system at the upper (lower) edge that will introduce a density modulation of the edge that, for the non-interacting system, will travel to the right (left). For the interacting system a density modulation introduced on the upper edge will instead separate into both a right-moving and a (smaller) left-moving component. Since the charge on each edge is separately conserved, the separation of the modulation into a right- and left-going  mode has to be accompanied by a further splitting of each of these into components on both edges. The situation is illustrated  qualitatively in Fig.1, where an initial wave packet of Gaussian form on the upper edge ($\chi=+$) dynamically splits into a dominant right-moving packet and a smaller left-moving one.  

\begin{figure}[h]
\begin{center}
\includegraphics[width=11cm]{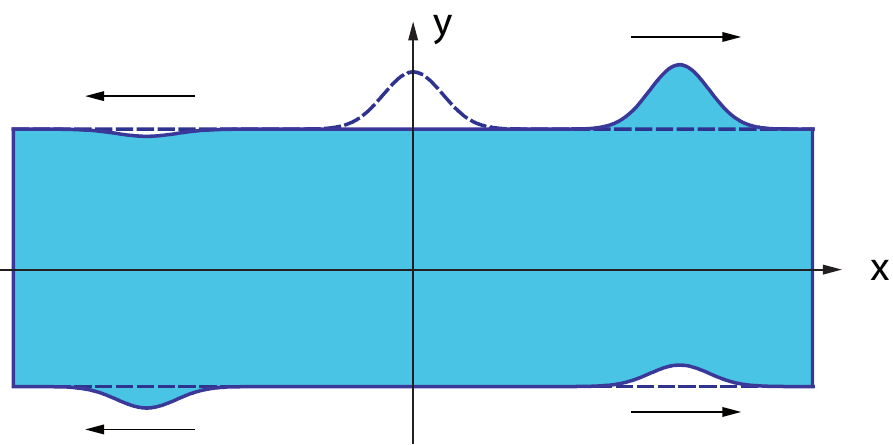}
\end{center}
\caption{\small Two-dimensional representation of chiral separation in an interacting system. A charge which is injected at the upper edge (with momentum close to  $k_F$, blue dashed curve) is separated into a a right-moving and a left-moving component. Each of these carries a density component both on the upper and on the lower edge. The ratio between the upper and lower part of the right-going charge is determined by the interaction parameter $g$, and is the same as the ratio between the lower and upper part of the left-going charge. The initial condition restricts the total charge on the lower edge to be zero. The figure is qualitative and for illustration purpose the interactions have been chosen rather strong with a parameter value about $g=0.5$. \label{Kiral}}
\end{figure}

The two-dimensional representation indicates that the injected charge is separated in {\em four} parts, which we may specify by the two quantum numbers $\chi$ and $\Gamma$. The first one ($\chi=\pm$) specifies the location on either the upper edge (at $+k_F$) or the lower edge (at $-k_F$), while the second quantum number $\Gamma=\pm$ distinguishes between the right-moving and left-moving components. For a well localized initial charge distribution, like the one illustrated in Fig.1 these components are all well defined.  There is thus a dynamical separation of the charge in the right- and left-going parts, and a frequency separation can further distinguish the parts that are close to $+k_F$ and $-k_F$. However, one should note that for a general charge distribution the total (integrated) charge has a clear separation only in the two conserved charges $N_\chi$ corresponding to particles on the upper/lower edge. There is in general no unique definition of the total right-moving or left-moving charge. This is an important point that we will now discuss further.

We first note that in the Fourier decomposition of the charge density the information about the total (global) charge sits in the $q=0$ component, while the local distribution of charge is determined by the $q\neq 0$ components. The periodicity conditions \pref{percond}, which restrict the charges $N_\chi$ of the upper and lower edges to integer values, only affect the $q=0$ component. On the other hand the separation in right- and left-going components is only meaningful for $q\neq 0$, since the $x$-independent $q=0$ component has no motion, neither to the right nor to the left. Therefore, the separation of the total charge $N$ into the two chiral components $Q_\pm$, as done in \pref{zerofields}, is in this sense somewhat arbitrary. In the bosonization of the theory the $q=0$ and the $q\neq 0$ components are treated in different ways, and the main point in the following will be to do that also in the discussion of the chiral separation of charge.

Following Ref.\onlinecite{HeinonenKohn87} we then introduce a distinction between the total {\em local} charge, defined as the $q\to 0$ component of the charge density, and the {\em global} charge, defined as the $q=0$ component.  For the two charges associated with the quantum number $\chi$ the local charges are denoted
\be{local}
\bar N_\chi= \lim_{q\to 0} \gr_{\chi q}
\ee
marked with a {\em bar} to indicate that they are not necessary equal to the global charges $N_\chi=\gr_{q=0}$. Note that the local charges, when defined in this way, do not refer to localization within a fixed line element $\gD x$. They should rather be viewed as the charge operators restricted to a line element $\gD x$, in the limit $\gD x\to \infty$, when this is taken {\em after}  the limit $L\to\infty$. (A more specific way of doing this will be discussed later.) We shall assume that the limit \pref{local} does exist for the states of interest. As opposed to the global charges $N_\chi$, the local charges $\bar N_\chi$ will take values that are not restricted by the boundary conditions. The difference between the two types of charges is due to the presence of a compensating background charge that affects only the $q=0$ mode. In the case of a compact space, which we consider here, this background charge will be evenly distributed over the circle. For an open space this compensating charge can instead be viewed as a boundary charge infinitely far away from the region of interest \cite{HeinonenKohn87}.

To find explicit expressions for the local {\em chiral} charges, we return to the expressions \pref{aop} for bosonic annihilation and creation operators, extract the charge density operators and re-express them in terms of the transformed Bose operators $b_q$ and $b_q^\dag$ (defined in \pref{bogo}). For $q\neq 0$ the operators are
\be{rhoq}
\gr_{\chi q}&=&\sqrt{\frac{L|q|}{2\pi}}
(\gq(\chi q)\,a_q+\theta(-\chi q)\,a_{-q}^\dag )\nn
&=&\sqrt{\frac{L|q|}{2\pi}}
(\gq(\chi q)\cosh\xi_q+\theta(-\chi q)\sinh\xi_q) \,b_q\nn
&&
+\sqrt{\frac{L|q|}{2\pi}}
(\gq(\chi q)\sinh\xi_q+\theta(-\chi q)\cosh\xi_q) \,b_{-q}^\dag
\ee
Since $b_q$ defines the positive frequency part of the operator, with time evolution $e^{-i\gw_q t}$, and $b_q^\dag$ defines the negative frequency part, with time evolution $e^{i\gw_q t}$, the time evolution of the operator $\gr_{\chi q}$ (for $q\neq 0$) is implicitly given by the expression \pref{rhoq}, and the right-and left-moving parts of the operator can therefore be extracted,
\be{fc}
\gr_{\chi \Gamma\, q}=\sqrt{\frac{L|q|}{2\pi}}
\left[\gq(\Gamma q)(\gq(\chi q)\cosh\xi_q+\theta(-\chi q)\sinh\xi_q) \,b_q\right. \nn
+\left. \gq(-\Gamma q)(\gq(\chi q)\sinh\xi_q+\theta(-\chi q)\cosh\xi_q) \,b_{-q}^\dag\right]
\ee
This gives the four-fold separation of the charge density specified by the two quantum numbers $\chi$ and $\Gamma$.

The local charges, which are defined by taking the limit $q\to 0$, we assume to be real valued, which means that the limit is independent of taking the limit $q\to 0^+$ or $q\to 0^-$. This further means that it is the hermitian part of the charge density \pref{fc} that is of interest. We write it as
\be{herm}
\half(\gr_{\chi \Gamma\, q}+\gr_{\chi \Gamma\, q}^\dag)
= [\gq(\Gamma\chi)\cosh\xi_q+\theta(-\Gamma\chi)\sinh\xi_q]\, B_{\gG|q|} 
\ee
with $B_q$ defined by
\be{Bq}
B_q=\half\sqrt{\frac{L|q|}{2\pi}}(b_q+b_q^\dag) \, .
\ee
We now take the limit $q\to 0$, and define
\be{limit}
B_\pm=\lim_{q\to 0^\pm} B_q \, .
\ee
This gives for the $\chi\Gamma$ components of the local charge
\be{charges}
\bar Q_{\chi\Gamma}=[\gq(\Gamma\chi)\cosh\xi_0+\theta(-\Gamma\chi)\sinh\xi_0] B_\Gamma \, ,
\ee
or written out separately
\be{sep}
&&\bar Q_{++}=\cosh\xi_0 B_+\,,\quad \bar Q_{-+}=\sinh\xi_0 B_+  \nn
&& \bar Q_{+-}=\sinh\xi_0 B_-\,,\quad \bar Q_{--}=\cosh\xi_0 B_-\,,
\ee
where the mixing parameters are determined by the Luttinger liquid parameter $g$ as
\be{par}
\cosh\xi_0=\frac{g+1}{2\sqrt{g}}\,,\quad \sinh\xi_0=\frac{g-1}{2\sqrt{g}} \, .
\ee
Since $B_\pm$ are time independent operators, all the four components of the local charge are separately conserved. The two $\chi$ components  of the local charge are furthermore given as
\be{sep2}
\bar N_+=\sum_\Gamma \bar Q_{+\Gamma}=\frac{1}{2\sqrt{g}}[(g+1)B_++(g-1) B_-]\nn
\bar N_-=\sum_\Gamma \bar Q_{-\Gamma}=\frac{1}{2\sqrt{g}}[(g-1)B_++(g+1) B_-]
\ee 
and the two chiral  components as
\be{sep3}
\bar Q_+=\sum_\chi \bar Q_{\chi +}=\sqrt{g} B_+\nn
\bar Q_-=\sum_\chi \bar Q_{\chi -}=\sqrt{g} B_-  \, .
\ee
From these expressions follow that, if we introduce local charges $\bar N=\sum_\chi\bar N_\chi\,,\; \bar J=\sum_\chi \chi \bar N_\chi$, we have the relation
\be{rel}
\bar Q_\pm=\half(\bar N\pm g \bar J)  \, ,
\ee 
which is of the same form as in the previous expression \pref{zerofields} used for the chiral charges. An important difference is, however,  that $\bar N$ and $\bar J$ are not, unlike $N$ and $J$, restricted to integer values. Therefore $\bar Q_\pm$ can also take more general values.

Note that even if the total local charge in \pref{sep} has been separated in four parts, these depend on only two charge operators, $B_\pm$. The operator $B_+$ is associated with the right-going component in such a way that there is, in the two-dimensional representation, a fixed ratio $\tanh\xi_0=(g-1)/(g+1)$ between the parts of this component on the upper and lower edges. Similarly $B_-$ determines the charge of the left-going component with the same ratio $\tanh\xi_0$ between the parts on the lower and upper edges. This means that for a right-moving component, the (small) charge moving on the lower edge can be regarded as a (reduced) image charge of the larger one moving on the upper edge. It is natural to view this as a polarization effect caused by the (long range) interaction $V_2$ which acts between the two edges.

In the above discussion we have assumed that the limit $q\to 0$ is well defined for the charge operators $B_q$. We are aware of the problem to make this assumption precise. Our intention, however, is to apply the charge operators only to states that  have well defined limits for the {\em expectation values }these operators. When we later discuss the charge fluctuations we have to be more careful and we will then be more specific about  how to take the limit.
\section{Examples}
We will now illustrate the general discussion by three examples. The first one is the case where an electron is injected at one of the Fermi points, and the charge of the electron splits into two non-integer parts which travel in opposite directions \cite{Pham00}. The second case corresponds to the situation where an initial state is created by injecting the electron at one Fermi point of the non-interacting system, and the interaction is then adiabatically turned on. In a Fermi liquid the corresponding procedure would turn an electron into a Landau quasi-particle of the interacting system. The electron charge is separated into a local charge and a background charge \cite{HeinonenKohn87} when the interaction is turned on, but in this case there is no counter-propagating charge component created. In the third example a local charge is created as a local polarization charge by applying an external field to the system. When the external potential is suddenly turned off the charge splits into two components which travel in opposite directions.

We derive in this section the expectation values of the local charge components in the three cases, and follow this up in the next section, where the charge fluctuations for the same three examples are examined.  Our discussion is based on the low-energy approximation to the interacting fermion system using the bosonized form. The limitations introduced by this approximation is assumed not to change results in any essential way, even if the charge fluctuations can not be regarded as a fully low-energy phenomenon. 
To check this, in the appendix  we compute  the charge fluctuations in the {\em non-interacting} fermion system
without making any low energy approximation, both for the ground state and for the  polarization charge induced by an external delta-function potential.

\subsection{Example 1: Sudden injection at a Fermi point}
We first consider the situation where an electron is injected on one of the edges, which means that the momentum of the particle is restricted to an interval which is close to either $+k_F$ or $-k_F$. This case is the one illustrated in Fig.\ref{Kiral}. The situation has been discussed in the paper by Pham et. al. \cite{Pham00} and we will compare our analysis of the chiral separation of the injected charge with theirs. The (global)  fermion numbers $N_\chi$ are in this case sharply defined, for example to be $N_+=1$ and $N_-=0$. When the particle is injected locally we expect no difference between the expectation value of the local and global charges, $\av{\bar N_\chi}=N_\chi$, since a local injection process cannot introduce a background charge that is evenly distributed around the whole circle.

With $\ket{G}$ as the ground state of the interacting many-fermion system, the state after injection is
\be{initial}
\ket{\Psi}=\Psi_\chi^\dag\ket{G} 
\ee
where $\Psi_\chi^\dag$ is a fermion creation operator that injects that particle at either the upper edge ($\chi=+$) or the lower one ($\chi=-$). Expressed in terms of the fermion field operator $\psi_\chi^\dag(x)$ it has the form
\be{creat}
\Psi_\chi^\dag=\int dx\, \phi(x)\psi_\chi^\dag(x)
\ee
where $\phi(x)$ is the wave function of the injected particle. The local charges are determined by the expectation value of the operator $B_q$, which we write as
\be{Bexp}
\bra G \Psi_\chi\, B_q\, \Psi_\chi^\dag \ket G
&=&\half\sqrt{\frac{L|q|}{2\pi}}\bra G \Psi_\chi\, (b_q+b_q^\dag)\, \Psi_\chi^\dag \ket G \nn
&=&\half\sqrt{\frac{L|q|}{2\pi}}(\bra G \Psi_\chi\, \com{b_q}{ \Psi_\chi^\dag} \ket G+\bra G \com{\Psi_\chi}{b_q^\dag} \Psi_\chi^\dag \ket G) \, ,
\ee  
where in the last expression we have used the fact that $b_q \ket G=0$.

Since $b_q$ and $b_q^\dag$ are directly related to the fermion density operator, they have  simple commutators with the fermion field operator
\be{bcom}
\com{b_q}{ \psi_\chi^\dag(x)} = \sqrt{\frac{2\pi}{L|q|}}[\gq(\chi q)\cq-\gq(-\chi q)\sq]e^{-iqx}\,\psi_\chi^\dag(x) \, .
\ee
This gives
\be{bcom2}
\com{b_q}{ \Psi_\chi^\dag} = \sqrt{\frac{2\pi}{L|q|}}[\gq(\chi q)\cq-\gq(-\chi q)\sq]\,\Psi_{\chi\,q}^\dag
\ee
where we introduced
\be{phiq}
\Psi_{\chi\,q}^\dag=\int dx\,e^{-iqx}\phi(x)\psi_\chi^\dag(x) \, .
\ee
For the expectation value of $B_q$ this gives
\be{Bexp2}
\bra G \Psi_\chi\, B_q\, \Psi_\chi^\dag \ket G
=\half[\gq(\chi q)\cq-\gq(-\chi q)\sq](\bra G \Psi_\chi\, \, \Psi_{\chi\,q}^\dag \ket G +\bra G \Psi_{\chi q}\, \, \Psi_\chi^\dag \ket G)
\ee
which determines the expectation value of $B_\pm$ in the limit $q\to 0$,
\be{Bpm}
\mean{B_\gG}=\lim_{q\to0^\gG}\bra G \Psi_\chi\, B_q\, \Psi_\chi^\dag \ket G
=\gq(\chi \gG)\cosh\xi_0-\gq(-\chi \gG)\sinh\xi_0  \, .
\ee
We here assumed $\lim_{q\to0^\pm}\Psi_{\chi\,q}=\Psi_\chi$ and used the normalization condition $\bra G \Psi_\chi\,  \Psi_\chi^\dag \ket G=1$. For $\chi=+\,$  this gives for the expectation values of the local charges,
\be{sepmean}
&&\mean{\bar Q_{++}}=\cosh^2\chi_0 \,,\quad\quad \mean{\bar Q_{-+}}=\so\co\,,\nn
&& \mean{\bar Q_{+-}}=-\sinh^2\chi_0 \,,\quad \mean{\bar Q_{--}}=-\so\co
\ee
and from this follows
\be{uplow}
\mean{\bar N_+} =1\,,\quad \mean{\bar N_-}=0 \, ,
\ee
which are the same as the corresponding values of the global fermion numbers $N_\pm$. For the chiral charges we find
\be{rightleft}
\mean{\bar Q_+} &=&\co(\co+\so)\quad=\half(1+g)\nn
\mean{\bar Q_-}&=&-\so(\co+\so)=\half(1-g)  \, .
\ee
If the fermion is instead injected at the lower edge, with $\chi=-$, the results are the same, if the signs of both $\chi$ and $\gG$ are changed.

The results \pref{rightleft} are the same as with the definition \pref{zerofields} for the (global) chiral charges $Q_\pm$, previously used in Ref. \onlinecite{Pham00}. This coincidence can be understood from the way the charge is introduced;  no constant background charge is created, and the limit $q\to0$ is therefore continuous. In the two next examples this will not be the case.
\begin{figure}[h]
\label{Kiral2}
\begin{center}
\includegraphics[width=11cm]{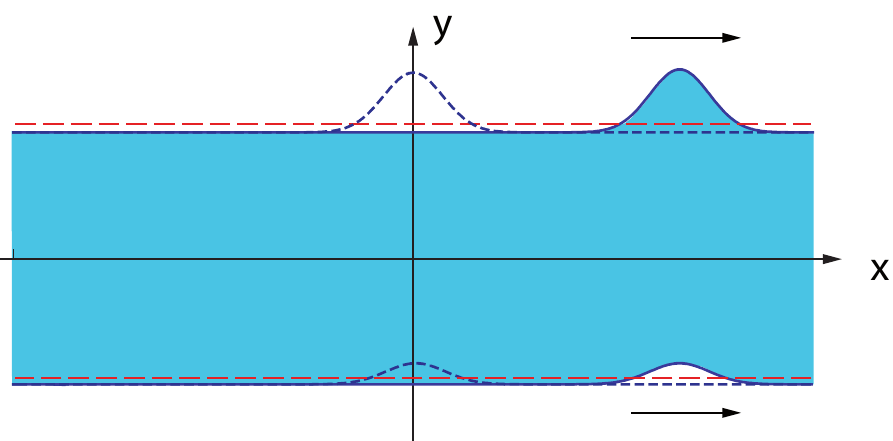}
\end{center}
\caption{\small Two-dimensional representation of the situation discussed in Example 2. A fermion is injected in the non-interacting system at the upper edge (\ie with positive chirality). The interaction is then adiabatically turned on to give a modified initial state indicated by the dashed blue curves. This is a positive chirality state which moves to the right during the time evolution of the interacting system. In this case no negative chirality component is created. However, constant background charges on the two edges compensate for the fact that the two local charges $\bar N_+$ and $\bar N_-$ are not identical to the conserved charges $N_+=1$ and $N_-=0$. The edge positions before injecting the particles are indicated by the dashed red lines.}
\end{figure}
\subsection{Example 2: Adiabatically turning on the interaction}
We next consider the situation where the particle is injected into the {\em non-interacting} system and the interaction is then adiabatically turned on. The energy eigenstates of the non-interacting system are then turned into eigenstates of the interacting system and chiral states of the non-interacting system are turned into ``dressed" chiral states of the interacting system. The initial state can now be written as
\be{initial2}
\ket{\Psi}=U \Psi_\chi^\dag\ket{F} 
\ee
where $\ket{F}$ is the ground state of the non-interacting system, $\Psi_\chi$ is the same fermion creation operator as in the previous example and $U$ is the unitary transformation from the non-interacting to the interacting system. We note that the initial state of the previous example can be written in a similar form, but with a change in the order of the operators, $\ket{\Psi}= \Psi_\chi^\dag U\ket{F}$.

We evaluate again the expectation value of the operator $B_q$,
\be{Bexp3}
\bra F \Psi_\chi\,U^\dag B_q U\, \Psi_\chi^\dag \ket F
&=&\half\sqrt{\frac{L|q|}{2\pi}}\bra F  \Psi_\chi U^\dag\, (b_q+b_q^\dag)\, U\Psi_\chi^\dag  \ket F\nn
&=& \half\sqrt{\frac{L|q|}{2\pi}}\bra F  \Psi_\chi \, (a_q+a_q^\dag)\, \Psi_\chi^\dag \ket F 
\ee
where we have used the relations
\be{abtrans}
a_q=U^\dag b_qU\,,\quad a_q^\dag=U^\dag b_q^\dag U
\ee
The expression has the same form as in the previous example, but now for the non-interacting case, corresponding to $\cosh\xi_q=1$ and $\sinh\xi_q=0$. This gives 
\be{Bexp5}
\bra F \Psi_\chi\,U^\dag B_q U\, \Psi_\chi^\dag \ket F
=\half \gq(\chi q)\bra{F}\Psi_\chi\Psi_{\chi q}^\dag+\Psi_{\chi q}\Psi_\chi^\dag\ket{F}
\ee
with the following value in the limit $q\to 0^\pm$,
\be{limval}
\mean {B_\pm}=\lim_{q\to 0^\pm}\bra F \Psi_\chi\,U^\dag B_q U\, \Psi_\chi^\dag \ket F
=\gq(\pm\chi)
\ee
For $\chi=+\,$ the local charges now are
\be{sepmean2}
&&\mean{\bar Q_{++}}=\co\mean {B_+}=\co \,,\quad
\mean{\bar Q_{-+}}=\so\mean {B_+}=\so\nn
&& \mean{\bar Q_{+-}}=\so\mean {B_-}=0 \,,\quad\quad\quad\quad
 \mean{\bar Q_{--}}=\co\mean {B_-}=0
\ee
which further gives
\be{uplow2}
\mean{\bar N_+} =\co=\frac{g+1}{2\sqrt{g}}\,,\quad \mean{\bar N_-}=\so=\frac{g-1}{2\sqrt{g}}
\ee
and
\be{rightleft2}
\mean{\bar Q_+} &=&\co+\so=\sqrt{g}\nn
\mean{\bar Q_-}&=& 0
\ee
We note that the adiabatic switching on of the interaction has transformed the state with positive chirality of the non-interacting system into a state with positive chirality of the interacting system. So in this case there is no splitting of the initial charge distribution into a right-moving and a left-moving component. Instead a constant background charge density has been created to account for the fact that local charge is not preserved under the adiabatic process. This gives the inequalities $\mean{\bar N_+}\neq N_+$ and $\mean{\bar N_-}\neq N_-$ which correspond to a discontinuous transition $q\to 0$. 

The results found here are consistent with those of Ref.~\onlinecite{HeinonenKohn87}, where a first order perturbative result is given for the local charge of the dressed fermion, with expansion in the deviation $g-1$ from the free theory.

\begin{figure}[h]
\begin{center}
\includegraphics[width=11cm]{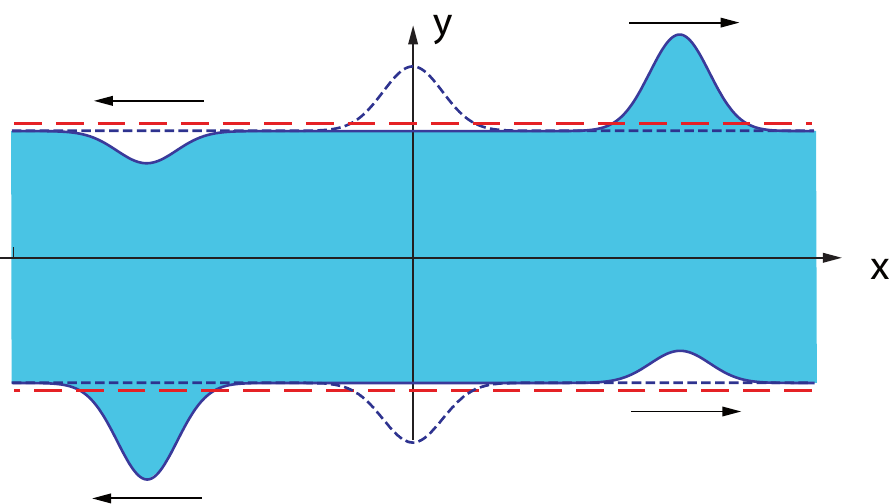}
\end{center}
\caption{\small Two-dimensional representation of the situation discussed in Example 3. A charge is initially attracted towards the point $x = 0$ by an external potential (dashed blue curve). The size of the local (polarization) charge  is determined by the strength of the potential. When the potential is next turned off the charge divides in two equal parts with different chiralities which move apart in opposite directions. The local charges are compensated by constant background charges of opposite signs. The edge positions before injecting the particles are in the figure indicated by the dashed red lines.\label{Kiral3}}
\end{figure}
\subsection{Example 3: Polarization charge}
As a third example we consider an initial state which is created by an external potential $W(x)$ that is slowly turned on. The potential  will attract a local charge. We assume in this case that no external charge is added, so that $N_+=N_-=0$. The initial state is the ground state of the Hamiltonian with the additional term,
\be{adham}
\gD H=\int dx \,W(x)\, \gr(x)={1\over L} \sum_q W_q\, \gr_{-q}=\frac{1}{L}N W_0+{1\over L} \sum_{\chi, q\neq 0} W_q\, \gr_{\chi-q}  \, ,
\ee
where $W_q$ are the Fourier components of the potential. The first term gives no contribution since $N=0$ and the second term we write in bosonic form as
\be{adham2}
\gD H&=&{1\over L} \sum_{q\neq 0} W_q\, \gr_{-q} \nn
&=&{1\over L} \sum_{q\neq 0}  \sqrt{\frac{L|q|}{2\pi}}
(\cosh\xi_q+\sinh\xi_q) (W_q^*\,b_q+W_q \,b_{q}^\dag)\nn
&\equiv& \sum_{q\neq 0}  (
\gD_q^*\,b_q+\gD_q\,b_q^\dag)
\ee
with
\be{delta}
\gD_q= \sqrt{\frac{|q|}{2\pi L}}
(\cosh\xi_q+\sinh\xi_q)W_q
\ee
The total (bosonized) Hamiltonian, in the subspace with $N=J=0$, then takes the form
\be{totham}
H'&=&\sum_{q\neq 0}(\hbar\gw_qb_q^\dag b_q+\gD_q^*\,b_q+\gD_q\,b_q^\dag)\nn
&=&\sum_{q\neq 0}\hbar\gw_q(b_q^\dag+\frac{\gD_q^*}{\hbar\gw_q}) (b_q+\frac{\gD_q}{\hbar\gw_q})-\sum_{q\neq 0}\frac{\gD_q^*\gD_q}{\hbar\gw_q}) \ee
where the last term is a constant that can be ignored. The first term has a form which is identical to the Hamiltonian without the external potential, except for a constant shift in the operators $b_q$ and $b_q^\dag$. This shift can be expressed as a unitary transformation
\be{utrans}
Sb_qS^\dag=b_q+\frac{\gD_q}{\hbar\gw_q}\,,\quad
Sb_q^\dag S^\dag=b_q^\dag+\frac{\gD_q^*}{\hbar\gw_q}
\ee
with
\be{utrans2}
S=\exp\left[\sum_q(\frac{\gD_q^*}{\hbar\gw_q}b_q-\frac{\gD_q}{\hbar\gw_q}b_q^\dag)\right] \, .
\ee
This gives $H'=S H S^\dag$ when the constant term in \pref{totham} is omitted, and the ground state of the Hamiltonian $H'$, with the external potential included, is therefore
\be{init}
\ket{\Psi}=\exp\left[\sum_q(\frac{\gD_q^*}{\hbar\gw_q}b_q-\frac{\gD_q}{\hbar\gw_q}b_q^\dag)\right]\,\ket{G} \, .
\ee
If the external potential is suddenly turned off, this state becomes the initial state which is then dynamically separated in a right-moving and a left-moving component. The situation is illustrated in Fig.\ref{Kiral3}. 

To find the local charges associated with the right-moving and left-moving components, we analyze the system in the same way as for the two previous examples. We find
\be{Bexp4}
\bra \Psi  B_q\,  \ket\Psi 
&=&\half\sqrt{\frac{L|q|}{2\pi}}\bra G S^\dag (b_q+b_q^\dag)\, S \ket G \nn
&=&-\half\sqrt{\frac{L|q|}{2\pi}}\frac{\gD_q+\gD_q^*}{\hbar\gw_q}\nn
&=& -\frac{|q|}{2\pi\hbar\gw_q} (\cosh\xi_q+\sinh\xi_q)\Re W_q
\ee
where we have used $b_q\ket G=0$. This gives for the limit $q\to 0$,
\be{Bpm2}
\av{B_\pm} =-\frac{W_0}{2\pi \hbar u}(\cosh\xi_0+\sinh\xi_0)=-\frac{W_0}{2\pi\hbar u}\sqrt{g}
\ee
with $W_0=\int dx W(x)$ and $u=\sqrt{v_J v_N}$ as the velocity of the chiral modes. We note that the polarization charge now separates in chiral components of equal value
\be{rightleft3}
\mean{\bar Q_+} = \mean{\bar Q_-} = -\frac{W_0}{2\pi \hbar u}g
\ee
The size of the charges thus depend on the strength of the external potential as well as on the parameters of the one-dimensional system. 
As a check on our calculation, we have verified that the result for the total charge found here is consistent with the expression for the full polarization tensor of the Tomonaga-Luttinger model which was obtained using perturbative methods\cite{Solyom79}. This also suggests 
that the phenomenon of chiral separation more generally could be understood as a dynamical polarization effect.

Also the local charges associated with the two values of the parameter $\chi$ have equal values,
\be{updown}
\mean{\bar N_\pm} = \mean{\bar Q_\pm} = -\frac{W_0}{2\pi \hbar u}g
\ee
and they are different from the global charges $N_\pm=0$.

\section{Charge fluctuations in the one-dimensional system}
An interesting and important question is to what extent the separation of a local charge in a right-moving and a left-moving component should be viewed as a separation of the charge itself or rather as a splitting of the probability amplitude, so that the charge moves either to the right or left without being split in two. The answer to this is not obvious when the local charge sits on the top of the background charge distribution of the many-particle ground state. In such a situation the charge fluctuations of the ground state tend to mask the sharpness of the additional local charge.

It is known from earlier discussions of the phenomenon of charge fractionalization in one-dimensional systems \cite{Kivelson82,Rajaraman82,GoldhaberKivelson91} that in some cases the effect of the background fluctuations can be filtered out by introducing a soft sampling function in the definition of the local charge.  This happens when there is a gap in the energy spectrum so that low frequency fluctuations are effectively suppressed by the gap while the soft edges of the sampling function makes it insensitive to high frequency fluctuations. The sharpness is then defined by the variance of the local charge collected by the sampling function over a finite region of the one-dimensional space.

In the present case the situation is different, since there is no gap in the low-energy spectrum. The background fluctuations therefore cannot be suppressed completely by defining a sufficiently soft sampling function. However, when sampling over a finite region the variance of the ground-state charge can be reduced to a finite value, and the variance of the additional local charge of the excited state can be measured relative to this value. In fact, for a system with a finite Fermi momentum $k_F$ the ground state fluctuations are finite (but large) even for a sampling function with sharp boundaries, and the effect of these can therefore in principle be subtracted. Our approach will therefore be to measure the fluctuations of the local charge relative to the ground-state fluctuations.  

The {\em local} charges, which we examine here, are always affected by the background fluctuations, while the {\em global} charges are not. Since  the (sharp) values of the global charges are fixed by the boundary conditions, they are insensitive to the dynamical splitting of the charge, and also to polarization effects.  Only for modified boundary conditions can the sharp charge values therefore be fractional. The important point is that the local charges are not restricted in any similar way. 

In the discussion to follow we first introduce a specific sampling function, which collects the local charge over a finite interval $a$.  We then use this to evaluate the ground-state fluctuations of the system and show that these diverge both in the infrared and in the ultraviolet. We next examine the charge fluctuations in the three examples discussed previously and show that for sufficiently large $a$ the fluctuations are identical to those of the ground state. As the last part of this section we relate the sharpness of the fractional charges in these cases to the fact that the initial state, in all the three cases, can be viewed as a coherent state of the bosonic variables in the long wave-length limit $a\to \infty$. For the case of the polarization charge, this is the case even for finite values of $a$.

We choose the following form for the sampling function 
\be{sampl}
f(x;a,b)=\half\left[{\rm erf}\left(\frac{x+a/2}{b}\right)-{\rm erf}\left(\frac{x-a/2}{b}\right)\right]  \, ,
\ee
which essentially equals $1$ for $-a/2<x<a/2$, and has an exponentially fast fall-off to $0$ outside this region, with $b$ as the characteristic length scale of the transition. The function $f(x;a,b)$ is well defined on the circle when $a<<L$, and has the Fourier transform 
\be{foursampl}
\tilde f_q(a,b)=\int_{-\infty}^{\infty} dx f(x;a,b)e^{-iqx}={2\over q}\sin(aq/2)e^{-b^2q^2/4} \, .
\ee
 The corresponding charge operator, for a given given chirality $\Gamma$, is
\be{qugamma}
 Q_{\Gamma}(a,b)&=&\int dx f(x;a,b)\gr_{\Gamma}(x)\nn
&=& {1\over L}\sum_q \tilde f_{-q}(a,b) \gr _{\Gamma q} \nn
&=&{4\over L} \sum_q\gq(\Gamma q)\frac{\sin(aq/2)}{q}e^{-{1\over 4}q^2b^2} (\cosh\xi_q+\sinh\xi_q)\,B_q \, .
\ee
In this expression we have disregarded the contribution from the $q=0$ term, which does not have a well defined separation in the two chiralities.  In the limit $L\to \infty$ with $a$ fixed, this term vanishes, and the expression for the charge operator is
\be{qugamma2}
 Q_{\Gamma}(a,b)={2\over \pi} \int_0^{\infty} dq \frac{\sin(aq/2)}{q}e^{-{1\over 4}q^2b^2} (\cosh\xi_q+\sinh\xi_q)\,B_{\Gamma q} \, .
\ee

We note that the expression earlier introduced for the (total) local charge $\bar Q_{\gG}$ is now recovered if, as the next step, we take the limit $a\to \infty$ with $b$ fixed, assuming a smooth transition $\lim_{q\to 0^\pm}B_q= B_\pm$, 
\be{qugamma3}
 \lim_{a\to\infty} Q_{\Gamma}(a,b)=(\cosh\xi_0+\sinh\xi_0)\,B_{\Gamma}={\bar Q}_\gG
\ee
This follows since large $a$ means that the integral only gets contributions from small $q$, with the following limit for the integral
\be{int}
\lim_{a\to\infty}  \int_0^{\infty} dq \frac{\sin(aq/2)}{q}e^{-{1\over 4}q^2b^2} ={\pi\over 2}
\ee
However, for the {\em variance} of the charge operator the limit $a\to\infty$ is not necessarily a smooth limit. In particular will the variance of the ground state charge diverge, and we examine therefore these fluctuations for finite $a$. For the excited states the charge fluctuations are finite when the ground-state contribution is subtracted, and with this in mind we shall still use the expression $\bar Q_\gG =\lim_{a\to\infty} Q_{\Gamma}(a,b)$ for the total local charge.
\subsection{Ground-state fluctuations}

For the ground state the expectation value of the charge operator \pref{qugamma2} vanishes since $b_q \ket G=0$ and therefore
\be{gs}
\bra G  B_q \ket G= \half\sqrt{\frac{L|q|}{2\pi}}\bra G b_q+b_q^\dag \ket G=0 \, .
\ee
The expression for the variance is, in the limit $L\to\infty$,
\be{gsfluct}
\gD  Q_{\Gamma}(a,b)^2&=&\bra G  Q^2_{\Gamma}(a,b) \ket G \nn
&=& {1\over \pi^2} \int_0^{\infty} dq \frac{\sin^2(aq/2)}{q}e^{-{1\over 2}q^2b^2} (\cosh\xi_q+\sinh\xi_q)^2\nn
&\approx& {g \over \pi^2} \int_0^{\infty} d\eta \frac{\sin^2\eta}{\eta}e^{-{2}\eta^2(b/a)^2} 
\ee
where in the last step we have assumed $a$ to be sufficiently large so that $\xi_q$ can be replaced by its $q=0$ value, with $g=(\cosh\xi_0+\sinh\xi_0)^2$. The integral can be solved in terms of a generalized hypergeometric function and gives
\be{gsfluct2}
\gD  Q_{\Gamma}(a,b)^2={ g\over 4\pi^2} {a^2\over b^2}\;_2F_2 (\{1, 1\}, \{3/2, 2\}, - a^2 / 2b^2)
\ee

\begin{figure}[h]
\begin{center}
\includegraphics[width=9cm]{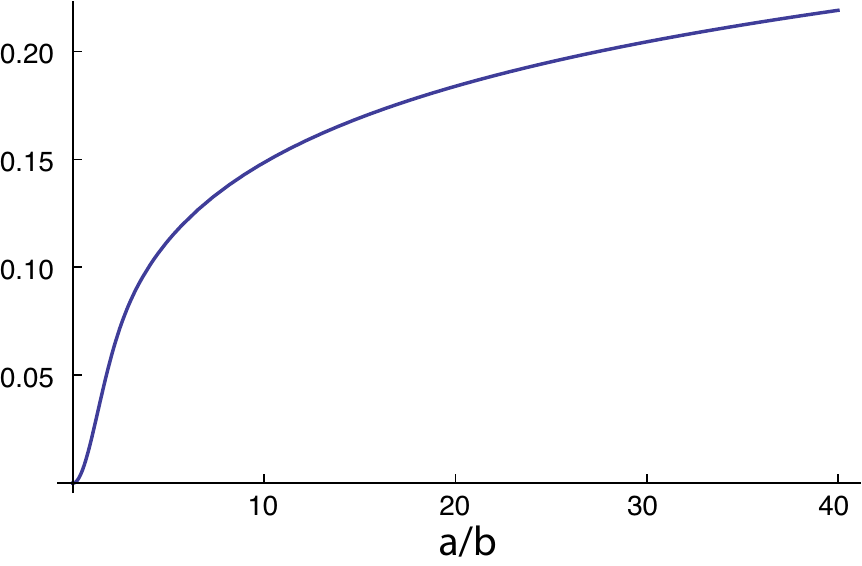}
\end{center}
\caption{\small Ground state fluctuations of the charge operators $Q_\gG(a,b)$. The plot shows the variance $\gD  Q_{\Gamma}(a,b)^2$ as a function of $a/b$ with the normalization $g=1$ for the interaction parameter. \label{Fluktuasjoner}}
\end{figure}

This result shows that $\gD  Q_{\Gamma}(a,b)^2$ is independent of the chirality $\gG=\pm$, as it should be, due to left-right symmetry. The fluctuations only depend on the ratio $a/b$, as follows from the scale invariance of the gapless theory,  and the functional form is shown in Fig.\ref{Fluktuasjoner}. Note that the interaction parameter $g$ only appears as a normalization factor. For large $a/b$ the function has a logarithmic behavior
\be{asymp}
\gD  Q_{\Gamma}(a,b)^2\approx {g\over 2\pi^2} \ln({a\over b})  \, .
\ee
The expression shows the expected logarithmic ultraviolet divergence when $b\to 0$, which corresponds to the situation where the sampling function approaches a step function. However, there is also   logarithmic infrared divergence when $a\to \infty$. In reality the ultraviolet behavior is an artifact of the approximation used in the boson representation. When the finite depth of the Fermi sea is taken into account that will introduce a physical cutoff, namely the Fermi momentum $k_F$. This cutoff essentially replaces $b$ with $1/k_F$ in the logarithm when $b\to 0$. The infrared divergence for large $a$ is however real, and  is due to the presence of massless excitations in the system.

\subsection{Charge separation and charge fluctuations} 
We consider next the three examples previously discussed, where a local charge is dynamically separated in a right-going and a left-going component. In all three cases we examine the fluctuations of the {\em total} local charge $\bar Q_\gG=\sqrt{g} B_\gG$ of each of the chiral components. 
Due to the presence of ground-state fluctuations we do not expect the fluctuations to vanish, even if the charges of each of the two chiral components in some sense are sharp. We shall here {\em define} the charges to be sharp if the fluctuations are identical to those of the ground state.
\vskip2mm
\noi
{\em Sudden injection}\\
The initial state in this case is $\ket{\Psi}=\Psi_\chi^\dag\ket{G}$ with $\Psi_\chi^\dag=\int dx\, \phi(x)\psi_\chi^\dag(x)$, where $\phi(x)$ is the wave function of the injected particle and $\chi$ specifies whether the particle is injected with momentum close to $k_F$ or $-k_F$ . Since the expectation value of the charge operator is already evaluated, we only need to determine the expectation value of the quadratic operator, and we examine therefore the expectation value of the operator $B_qB_{q'}$,
\be{expbq}
&&\bra G \Psi_\chi  B_qB_{q'} \Psi_\chi^\dag\ket G={{L\sqrt{|q||q'|}}\over{8\pi}} 
\bra G \Psi_\chi  (b_q b_{q'}+b_q^\dag b_{q' }+
b_{q'}^\dag b_q +b_q^{\dag}b_{q'}^{\dag} 
+ \gd_{qq'}) \Psi_\chi^\dag\ket G
\nn
&&={{L|q|}\over{8\pi}}\gd_{qq'} +{{L\sqrt{|q||q'|}}\over{8\pi}}  \left\{\bra G \Psi_\chi  \com{b_q}{\com{b_{q'}}{\Psi_\chi^\dag}}\ket G\right.
\left.+ \bra G \com{\Psi_\chi}  {b_q^\dag}\com{b_{q'}}{\Psi_\chi^\dag}\ket G\right.\nn
&&\quad\quad+ \bra G \com{\Psi_\chi}  {b_{q'}^\dag}\com{b_{q}}{\Psi_\chi^\dag}\ket G+ \left.\bra G \com{\com{\Psi_\chi}  {b_q^\dag}}{b_{q'}^\dag}\Psi_\chi^\dag\ket G\right\}
\ee 
Here we again have used that $b_q$ annihilates the ground state. The commutators in this expression can be evaluated by use of \pref{bcom2}  to give 
\be{expq}
\bra G \Psi_\chi  B_q B_{q'} \Psi_\chi^\dag\ket G&=&{{L|q|}\over{8\pi}}\gd_{qq'} 
+{1\over 4}(\gq(\chi q)\cosh\xi_q-\gq(-\chi q)\sinh\xi_q)\nn
&&\times(\gq(\chi q')\cosh\xi_{q'}-\gq(-\chi q')\sinh\xi_{q'})\nn
&&\times\left\{\bra G   \Psi_\chi\Psi_{\chi\,(q+q')}^\dag \ket G
+ \bra G   \Psi_{\chi\,q}\Psi_{\chi\,q'}^\dag \ket G
+ c.c.\right\}
\ee
where $\Psi_{\chi\,q}$ is defined by \pref{phiq}. The first term in this expression is identical to the ground-state expectation value of the operator. In the limit $L\to\infty$ the expectation value of the squared charge operator then has the form 
\be{expq2}
&&\bra \Psi   Q_{\Gamma}(a,b)^2 \ket \Psi 
={1\over \pi^2} \int_0^{\infty} dq \frac{\sin^2(aq/2)}{q}e^{-{1\over 2}q^2b^2} (\cosh\xi_q+\sinh\xi_q)^2 \nn
&&\quad\quad+ {1\over \pi^2}\int_0^\infty dq \int_0^\infty dq' \frac{\sin(aq/2)}{q} \frac{\sin(aq'/2)}{q'}
e^{-{1\over 4} b^2(q^2+q'^2)}\nn
&&\quad\quad\quad \times\,(\gq(\chi\gG)\cosh\xi_q\cosh\xi_{q'}+\gq(-\chi\gG)\sinh\xi_q\sinh\xi_{q'})\nn
&&\quad\quad\quad\times\,(\cosh\xi_q+\sinh\xi_q)(\cosh\xi_{q'}+\sinh\xi_{q'})\nn
&&\quad\quad\quad\times\,\left\{\bra G   \Psi_\chi\Psi_{\chi\,(\gG q+\gG q')}^\dag \ket G
+ \bra G   \Psi_{\chi\,(\gG q)}\Psi_{\chi\,(\gG q')}^\dag \ket G
+ c.c.\right\}
\ee
One should note the different behavior of the ground state contribution, on the first line, and that of the remaining contribution when the limit $a\to \infty$ is taken. In the first term the oscillating factor does not give an effective limitation in the contribution to the q-integral for large $q$. The integral is instead limited by the exponential factor, so that effectively one has $q\lesssim1/b$. Furthermore the $1/q$ factor gives rise to the logarithmic dependence of $a/b$. The oscillating factors in the second term instead introduces the effective limitations $q,q'\lesssim1/a$. For the integrand the large $a$ limit can therefore be interpreted as $q,q'\to0$, and this gives a finite contribution when $a\to\infty$. The expectation value in this limit therefore simplifies to
\be{expq3}
 \mean{\bar Q_{\Gamma}^2}&=&\mean{ \bar Q_{\Gamma}^2}_0+(\gq(\chi\gG)\cosh^2\xi_0+\gq(-\chi\gG)\sinh^2\xi_0)(\cosh\xi_0+\sinh\xi_0)^2\nn
 &=&\mean{ \bar Q_{\Gamma}^2}_0+ \mean{\bar Q_{\Gamma}}^2
\ee
where the term labeled $0$ is the divergent ground state contribution. This expression shows that the charge fluctuations of the state $\ket{\Psi}=\Psi_\chi^\dag\ket{G}$, in the limit $a\to\infty$, are identical to those of the ground state,
\be{chfluct}
\gD  \bar Q_{\Gamma}^2=(\gD  \bar Q_{\Gamma})_0^2 \, .
\ee
This result we expect to be true not only when $a$ tend to infinity but when $a$ is much larger than the width of the wave function $\gf(x)$ of the injected particle.

We conclude that when a fermion is injected with a sharp value for $\chi$, it will spontaneously split in two parts of different chirality and with non-integer charges $\half(1\pm g)$, where each of the charges is sharp in the meaning that the charge fluctuations are identical to those in the ground state.

\vskip2mm
\noi
{\em Adiabatic initial state}\\
The initial state in this case is $\ket{\Psi}=U \Psi_\chi^\dag\ket{F}$ with $\ket{F}$ as the non-interacting ground state and $U$ as the unitary transformation to the interacting system. As discussed in Sect.4.2, the effect of this transformation is to replace the matrix elements of $B_q$ for the interacting system with the corresponding ones for the non-interacting system. In the present case this means that \pref{expq} is replaced by
\be{nonin}
\bra F \Psi_\chi  U^\dag B_q B_{q'} U  \Psi_\chi^\dag\ket F&=&{{L|q|}\over{8\pi}}\gd_{qq'} 
+{1\over 4}\gq(\chi q)\gq(\chi q')\nn
&&\times\left\{\bra F   \Psi_\chi\Psi_{\chi\,(q+q')}^\dag \ket F
+ \bra F   \Psi_{\chi\,q}\Psi_{\chi\,q'}^\dag \ket F
+ c.c.\right\}
\ee
The corresponding expression for the expectation value of $Q_\gG^2(a,b)$ gives in the limit $a\to\infty$ 
\be{expq4}
 \mean{\bar Q_{\Gamma}^2}&=&\mean{ \bar Q_{\Gamma}^2}_0+\gq(\chi\gG)(\cosh\xi_0+\sinh\xi_0)^2\nn
 &=&\mean{ \bar Q_{\Gamma}^2}_0+ \mean{\bar Q_{\Gamma}}^2
\ee
from which the same equality follows as in the previous example,
\be{chfl2}
\gD  \bar Q_{\Gamma}^2=(\gD  \bar Q_{\Gamma})_0^2
\ee
In the present case the initial condition gives rise to a propagating wave with only one chirality, which carries the charge $\sqrt{g}$. It is also here sharp in the sense that the fluctuations are identical to those in the ground state.
\vskip2mm
\noi
{\em Polarization charge}\\
This is the case where a local charge is created not by injecting a particle from the outside but by polarizing the one-dimensional system with an external potential. The initial state is now  $\ket{\Psi}=\exp\left[\sum_q(\frac{\gD_q^*}{\hbar\gw_q}b_q-\frac{\gD_q}{\hbar\gw_q}b_q^\dag)\right]\,\ket{G}$, with the coefficient $\gD_q$ defined by \pref{delta}. The state is in fact a {\em coherent state}, since the effect of the unitary transformation that maps from $\ket{G}$ to $\ket{\Psi}$ is simply to introduce a c-number addition to the operators $b_q$ and $b_q^\dag$, as shown by \pref{utrans}. This means that for any linear combination of $b$ and $b^\dag$ operators the fluctuations are the same as in the ground state. This is in particular so for the operators $B_q$ and for a product of two operators we have the identity
\be{fact}
\bra\Psi B_qB_{q'}\ket\Psi-\bra\Psi B_q\ket\Psi\bra\Psi B_{q'}\ket\Psi=
\bra G B_qB_{q'} \ket G-\bra G B_q\ket G\bra G B_{q'}\ket G
\ee
For the charge fluctuations this implies
\be{fluct}
\gD Q_\gG^2(a,b)=(\gD Q_\gG^2(a,b))_0
\ee

The conclusion is that also here the fractional charges associated with the two chiral components are sharp in the sense that the fluctuations are identical to those in the ground state. 
However, one should note that the fluctuations in this case are identical with those in the ground state, not only for large $a$, but for any $a$ and $b$, even if this means that the sampling function catches only a fraction of the full charge distribution. 

It is of interest to note that, as opposed to the previous two examples, the chiral charges $\mean{\bar Q_\pm}$ in this case may be fractional even in the non-interacting case (\ie for  $g=1$). This is so, since the value of the charges depends not only on $g$ but also on the strength of the external potential, as shown by \pref{updown}. That chiral separation of sharp fractional charges is not an exclusive property of the Luttinger liquid, but occurs also in the free Fermi gas, emphasizes that it naturally can be seen as a polarization phenomenon. 
In the appendix we make use of this by examining the polarization effect for non-interacting fermions without applying the low energy
approximation of the boson representation of the theory. A comparison of the results here and of those in the appendix is of interest in order to check explicitly that the high frequency contributions to the fluctuations do not in any essential way change the results of this section.

\subsection{Coherent states and fluctuations}
The sharpness of the polarization charges discussed above can be ascribed to that the state created by the external potential is, in the low energy approximation, a coherent state in the boson variables. It is in fact a coherent state in all the bosonic modes labeled by the frequency variable $q$,  and this is the reason that not only the integrated local charge but {\em even the charge density is a sharp quantum observable}. However, also in the two other cases we have examined the initial state can, in an approximate sense, be regarded as a coherent state. This is not the case for any $q$ component, but it is true in the limit $q\to 0$, which can be seen as a reason for the sharpness of the  total local charge, while in these cases the charge density is not sharp in the same sense. We shall discuss this point a bit further, first for the case of a sudden injection of a fermion at one of the Fermi points.

We consider then the action of $b_q$ on the initial state $\ket\psi=\Psi_\chi^\dag\ket G$,
\be{bq}
b_q \Psi_\chi^\dag\ket G=\com{b_q }{\Psi_\chi^\dag}\ket G
\ee
where $b_q$ annihilates the ground state $\ket G$. The commutator has previously been given in Eq.\pref{bcom}, and in the limit $q\to 0$ this gives
\be{bqlim}
b_q \Psi_\chi^\dag\ket G\to \gb_{\pm\chi} \Psi_\chi^\dag\ket G\,,\quad q\to0^\pm
\ee
with
\be{beta}
\gb_\chi=\sqrt{\frac{2\pi}{L|q|}}[\gq(\chi )\cosh\xi_0-\gq(-\chi )\sinh\xi_0]
\ee
Eq.~\pref{bqlim} shows that the initial state $\ket\Psi$ is an eigenstate of $b_q$, and therefore (approximately) a coherent state, for sufficiently small $q$. This implies that observables that are linear combinations of $b_q$ and $b_q^\dag$ operators for sufficiently small $q$, will have quantum fluctuations in this state that are equal to those in the ground state. This is in particular true for the total local charge operator $\bar Q_\gG$.

In the case where the interaction is adiabatically turned on the initial state is $\ket\Psi=U \Psi_\chi^\dag\ket F$, where $\ket F$ is the ground state of the non-interacting system and $U$ is the unitary operator that transforms from the Hamiltonian of the non-interacting system into the Hamiltonian of the interacting system. In this case we have
\be{bq2}
b_q \,U \Psi_\chi^\dag\ket F=U a_q \Psi_\chi^\dag\ket F\to \ga_{\pm\chi}\, U\Psi_\chi^\dag\ket G\,,\quad q\to0^\pm
\ee
where the eigenvalue $\ga_{\pm\chi}$ is the non-interacting ($g=1$) version of the coefficient $\gb_{\pm\chi}$,
\be{alpha}
\ga_\chi=\sqrt{\frac{2\pi}{L|q|}}\gq(\chi )
\ee
Since the initial state $\ket\Psi$ can be regarded as an eigenstate of the bosonic annihilation operators $b_q$ for sufficiently small $q$, the situation is the same as for sudden injection of the fermion. In both cases the form of the initial state as a coherent state for small $q$ can be seen as an explanation for the fact that the fluctuations of the total local charges are identical to those of the ground state.

In both cases discussed above, with a sudden or adiabatic injection of an additional particle, the wave function of the particle has been assumed to have a well defined value of the quantum number $\chi$. This means that the particle is injected with momentum close to one of the Fermi points $k_F$ or $-k_F$. As a contrast to these situations let us now consider the case where a fermion is instead injected as a superposition of components with $\chi=+$ and $\chi=-$. This corresponds to replace, in the above expressions, the fermion creation operator $\Psi_\chi$ with a composite operator of the form $\Psi^\dag=\sum_\chi c_\chi \Psi^\dag_\chi$. The corresponding states $\Psi^\dag\ket G$ and $U \Psi^\dag\ket F$ then are no longer coherent states in the sense expressed by \pref{beta} and 
\pref{alpha}. This follows since the coefficients $\gb_\chi$ and $\ga_\chi$ have different 
values for $\chi=+$ and $\chi=-$.  The variance of the chiral charge in both cases can be written as 
\be{chvar}
\gD\bar Q_\gG^2=(\gD\bar Q_\gG)_0^2+\sum_\chi(|c_\chi|^2-|c_\chi|^4)\mean{\bar Q_\gG}_\chi^2-2|c_+|^2|c_-|^2 \mean{\bar Q_\gG}_+\mean{\bar Q_\gG}_-
\ee
with
$\mean{\bar Q_\gG}_\chi=\bra{G}\Psi_\chi \bar Q_\gG\Psi^\dag_\chi \ket G$
in the case of sudden injection and 
$\mean{\bar Q_\gG}_\chi=\bra{F}\Psi_\chi U^\dag \bar Q_\gG U \Psi^\dag_\chi \ket F$ in the adiabatic case.
Unless one of the coefficients $c_\chi$ has absolute value $1$ and the other is zero, there will now be non-vanishing deviations in the variance from the ground state value, and the mean value of the observable $\bar Q_\gG$ in the state $\ket\Psi=\Psi^\dag\ket G$ does therefore not correspond to a quantum mechanically sharp value. The chiral separation of the initial charge in this case should therefore not be seen as a splitting of the charge itself but rather of the probability for the charge to move to the right or to the left.

In fact, it is clear from general reasoning that any state which has quantum mechanically sharp values for the charges $\bar Q_\gG$ (as well as for $\bar N_\chi$) has to be characterized by well-defined, quantized values for the global charges $N_\chi$. The reason for this is the following. The local charges $\bar Q_\gG$ commute with the global charges $N_\chi$. This implies that the matrix elements of  $\bar Q_\gG$ as well as of $\bar Q_\gG^2$ vanish between states that have different quantized values for either $N_+$ or $N_-$. Consequently, for a state $\ket\Psi$ which is a superposition of such states, the expectation values of  $\bar Q_\gG$ and $\bar Q_\gG^2$ will only depend on the {\em absolute square} of the expansion coefficients of $\ket\Psi$ in eigenstates of $N_\chi$ with different eigenvalues. Eq. \pref{chvar} shows this for the particular case where a single particle is added to the system. Also the variance $\gD\bar Q_\gG^2$ will therefore only depend on the absolute square of the expansion coefficients, which means that it has the same form as for a {\em statistical mixture} of the same eigenstates. This implies that the charges $\bar Q_\gG$, for such a superposition are not quantum mechanically sharp (unless the mean value of the operators are precisely the same for the components with different values of $N_\chi$).

This makes it possible to illustrate in a simple way the difference between the two situations where, on one hand chiral separation gives rise to a separation of the probability for the charge to move in one direction or the other, and on the other hand the situation where the charge is split in two sharply defined smaller charges with opposite chiralities. We may consider the case of non-interacting fermions, where the first case corresponds to making a sudden injection of a particle in a superposition of the two chiralities. The charge is then not split in two, but the particle moves with a non-vanishing probability for each of the two chiralities. In the other case a polarization charge is created by an external potential. When the charge is released by turning off the potential it divides in two chiral components, and both of these are sharp in the meaning that the charge fluctuations are identical to those of the ground state. The expectation values $\mean {\bar Q_\gG}$ may be equal in theses two cases, but the fluctuations $\gD\bar Q_\gG^2$ will be different.

\section{Summary and outlook}
In this paper we used the concept of {\em local charge} to  discuss the character of the chiral separation of charges that takes place in one-dimensional Luttinger liquids. Depending on the initial conditions the expectation values for the chiral components of the charges may take different, non-integer values, and we have examined to what extent these charges can be regarded as quantum mechanically sharp. Since the system has a gapless spectrum the question of quantum sharpness of an observable is less clear than for systems with a gapped spectrum. We have here defined {\em sharp} to mean that the quantum fluctuations of an observable in a given state are identical to the fluctuations of the observable in the many-particle ground state.

We have examined the mean values and fluctuations of the chiral charge components with three different initial conditions for the quantum state. The first one corresponds to a sudden injection of a fermion at one of the Fermi points, the second one to injection of a particle in the non-interacting system followed by an adiabatic switching on of the interaction, and the third one corresponds to creation of a local polarization charge by an external potential.   In all three cases the charges of the two chiral components of the quantum state have been found to have non-integral values  and to be sharp in the sense that the charge fluctuations are identical to those of the ground state. In particular we have found that sharp chiral charges of {\em arbitrary} values can be created as polarization charges by applying an external potential. This is true also for the non-interacting Fermi gas which shows that chiral separation is a polarization phenomenon which is not exclusive to Luttinger liquids. 

The quantum state created by the external potential has been shown to be a coherent state in the bosonic variables. Also in the two other cases the quantum state is a coherent state in an approximate sense, restricted to the bosonic variables of low frequency. This property, we believe, provides  an explanation for the charge fluctuations to be identical to the ground state fluctuations. However, since the representation of the states as coherent states is linked to the low-frequency bosonized formulation of the theory, this raises the question if the same conclusion concerning the sharpness of the non-integer charges can be drawn for the fermion system with a finite Fermi momentum, where the high frequency contributions are different. The comparisons we have made in the appendex for the case of non-interacting fermions indicate that the high frequency effects will in practice not change the conclusion concerning the sharpness of the non-integer charges. 

A related question concerns the importance of the one-dimensionality for the results. The non-integer charges studied here can be viewed as arising as due to polarization effects in the many-fermion system. Clearly there are related effects also in higher dimensions, and the question is whether non-integer charges which are quantum mechanically sharp, in the meaning used here, can be present also there. This problem has been addressed for systems with a gap\cite{GoldhaberKivelson91}, but we know of no analysis of the gapless case. The close link to the coherent states of the boson representation may indeed suggest that one-dimensionality is important for the effect, but  apart from this observation we will have to leave this as an interesting open question.

Although the examples we have analyzed show that quantum mechanically sharp charges of any value can in principle be created, this does not mean that sharp quantum values is a typical feature for the local charges of arbitrary quantum states. In the examples we have discussed the quantum states are characterized by quantum well-defined, sharp values for the {\em global} fermion numbers $N_\chi$. In particular, in the two first examples, a single fermion is added with well defined value of the quantum number $\chi$. If this particle is instead injected as a superposition of components with both $\chi=+$ and $\chi=-$ the local charges of the right-and left-going modes will no longer be sharp, but will have charge fluctuations of the same form as for a statistical mixture of the two components with different values of $\chi$. 

In this paper we have not discussed how to actually observe the fractional charges in 
question, but this most important question has been addressed in some recent papers. In 
the experiment described in Refs.  \onlinecite{Steinberg08} and \onlinecite{LeHur08},  electrons are injected by tunneling into a quantum 
wire at one of the Fermi points, and the chiral charge fractionalization is determined from 
a left-right asymmetry in the resulting current. 
The experimental result is consistent with 
the prediction of Ref.  \onlinecite{Pham00}  and thus also with our result \pref{rightleft} for the value of the local chiral 
charges. (We are aware that a different conclusion concerning this
experiment is reached in Ref. \onlinecite{Pugnetti08}.) It is not clear to us that this kind of measurement can be used to infer the presence of {\em sharp} local charges, but we note that the conclusion in Ref.~\onlinecite{LeHur08} about  sharpness of the charge is the same as reached here, there based on a calculation of the zero frequency noise in the current.

Noise measurements is also the subject of a recent  theoretical paper by  Berg \etal which considers the detection of fractional charges in a physical 
system very similar to the two dimensional model discussed in this paper \cite{Berg08}. The difference is that the width of the Hall bar is assumed to change continuously with a constriction in the middle. Away from the constriction the system is essentially free, with $g = 1$, while  $g < 0 $ in the middle region. A proposed experiment is to inject electrons on the upper edge via a contact, and measure the absolute value and the noise of the reflected current on the lower edge. The result of the calculation is consistent with thinking of the process as a reflection at a sharp boundary  between a lead with $g=1$ and a wire with $g < 0 $.
Although the situation is similar to the one discussed in some detail in this paper, where a unit charge inserted at the upper edge spontaneously separates in a right-moving and a left-moving part, the presence of a boundary makes it different.  In particular the value of the left-moving charge on the lower edge is not the same in the two cases.  
It would be of interest to extend our analysis to this case, with reflection of a fractional charge at a sharp boundary, and in particular to evaluate the fluctuations in the reflected charge. However, as far as the suggested experiment is concerned  it is not obvious to us that reflection towards a sharp boundary is the best way to model  situation. For a sufficiently smooth constriction, we think that our example 2, i.e., a adiabatic switching of the interaction, might in fact be more appropriate.

The results found in this paper, that states with arbitrary, sharp values for the local charges of the two chiral components  can (in principle) be created,  seem to cast some doubt on ideas that have been proposed, that quasi-particles with specific (fractional) values for the fermion numbers are the natural charge carriers of the Luttinger liquid \cite{Pham00}. However, we believe that the idea that fractionally charged quasi-particles have a natural place in the description of Luttinger liquids may not be ruled out. In particular, when we consider a system of non-interacting fermions, there are low-energy   
excitations which carry charges with non-integer, sharp values. Even so, the fundamental fermions, which carry unit charges, play a special role in the low-energy description, since they define a system of {\em free} particles. 

The idea that this picture can be generalized to the interacting case, where the low energy theory can be described in terms of non-interacting quasi-particles characterized by fractional charge and {\em fractional statistics} is an attractive one. We refer the reader to the interesting ideas of Isakov \cite{Isakov} and Wu \cite{Wu00} which relate the Luttinger liquid to systems of particles with fractional exclusion statistics.  It is also interesting to note that Heinonen and Kohn, in the paper we have referred to \cite{HeinonenKohn87}, assume that the fractional charge, calculated in the situation where the interaction is adiabatically turned on, is identical to the (local) charge carried by the Landau quasi-particles, and in that context they do not make a distinction between quasi-particles in one and higher dimensions. The questions concerning the meaning of such quasi-particles with fractional charge in Luttinger liquids remain as interesting questions for future investigation, and this goes beyond the discussion of how to characterize these charges as being quantum mechanically sharp or not, which has been the main aim of this paper.

\appendix

\section{Fluctuations in the free fermion system}
In this appendix, we examine the charge fluctuations of the non-interacting fermion system, first for the filled Fermi sea, \ie with the system in the ground state, and then in the presence of a delta function potential. In the latter case we also calculate the mean value of the induced charge.

\subsection{Ground state fluctuations}
Rather than using the  boson representation of the theory we apply the full many-particle description. The aim is to see explicitly to what extent high frequency fluctuations lead to different results in the two approaches. We note that, compared to the bosonized description, there is an additional frequency parameter present in the many-particle approach, namely the Fermi momentum $k_F$. It provides a high frequency cutoff, and it is of interest to see precisely how it appears in the expressions. The high frequency cutoff $b$, introduced earlier as a smoothness parameter of the sampling function $f(x;a,b)$, can now be set to zero giving a  sampling function with sharp steps, 
\be{ffunc}
f(x;a)=
\left\{ 
\begin{matrix}
\;\;1 &\quad -a/2<x<a/2 \cr
\;\;0 &\quad |x|>a/2 \ \ \ \ .
\end{matrix} 
\right. 
\ee
The corresponding charge operator is
\be{chop}
Q_a=\int dx f(x;a) \psi^\dag(x) \psi(x)=\frac{1}{L}\sum_{k,k'} \tilde f(k-k',a) c_k^\dag c_{k'}
\ee
where the Fourier transform of the sampling function is given by
\be{four}
\tilde f(k;a)=\int dx e^{-ikx}f(x;a)=2\,\frac{\sin ka/2}{k} \, .
\ee
The ground state expectation value of the operator is
\be{expval}
\mean {Q_a}=\frac{1}{L}\sum_{k=-k_F}^{k_F} \tilde f(0;a)=\frac{a}{L} N_0  \, ,
\ee
the quadratic charge operator is 
\be{quad}
Q_a^2=\frac{1}{L^2}\sum_{k_1,k'_1}\sum_{k_2,k_2'} \tilde f(k_1-k_1',a)\tilde f(k_2-k'_2,a) c_{k_1}^\dag c_{k_1'}c_{k_2}^\dag c_{k_2'} \, ,
\ee
and the ground state expectation value of the operator is
\be{quadexp}
\mean{Q_a^2}&=&\frac{1}{L^2}\sum_{k_1=-k_F}^{k_F}\sum_{k_2=-k_F}^{k_F}\tilde f(0;a)^2
-\frac{1}{L^2}\sum_{k_1=-k_F}^{k_F}\sum_{k_2=-k_F}^{k_F}\tilde f(k_1-k_2;a)^2\nn
&&+\frac{1}{L^2}\sum_{k_1=-k_F}^{k_F}\sum_{k_2=-\infty}^{\infty}\tilde f(k_1-k_2;a)^2\nn
&=& \mean{Q_a}^2 +\mean{Q_a}-\frac{1}{L^2} \int_{-a/2}^{a/2} dx \int_{-a/2}^{a/2} dx'\left( 
\sum_{k=-k_F}^{k_F}e^{-ik(x-x')}\right)^2 \, .
\ee
For the charge fluctuations this gives
\be{chfluct2}
\gD Q_a^2=N_0\frac{a}{L}-\frac{1}{L^2} \int_{-a/2}^{a/2} dx \int_{-a/2}^{a/2} dx'\frac{\sin^2(\pi N_0\frac{x-x'}{L})}{\sin^2(\pi \frac{x-x'}{L})} \, .
\ee
By exploiting the symmetries of the integrand and changing to new variables $X=\half(x+x')$ and $\xi=x-x'$, the expression can be re-written as
\be{chfluct3}
\gD Q_a^2=N_0\frac{a}{L}-\frac{2}{L^2} \int_{0}^{a} d\xi (a-\xi)\frac{\sin^2(\pi N_0\frac{\xi}{L})}{\sin^2(\pi \frac{\xi}{L})} \, .
\ee
We take the limit $a/L \to 0$ with $N_0/L = k_F/\pi$ kept constant, and further introduce a new integration variable by $¢\xi=a\eta$ and a new parameter $\gb=\pi a N_0/L= a k_F$. This gives
\be{chfluct4}
\gD Q_a^2=\frac{1}{\pi}\left[\gb-\frac{2}{\pi} \int_{0}^{1} d\eta (1-\eta)\frac{\sin^2\gb\eta}{\eta^2}\right] \, .
\ee
The integral can be expressed in terms of special functions as
\be{chfluct5}
\gD Q_a^2&=&\frac{1}{\pi^2}\left[1+\gg+\pi\gb-\cos2\gb-{\rm Ci}(2\gb)-2\gb\, {\rm Si}(2\gb)+\ln2\gb\right]\nn
&=&\frac{1}{\pi^2}\left[1+\gg+\pi ak_F-\cos2ak_F-{\rm Ci}(2ak_F)-2ak_F\, {\rm Si}(2ak_F)+\ln2ak_F\right] \, ,
\ee
where $\gg$ denotes Euler's constant, Ci the cosine integral function and Si  the sine integral function.

\begin{figure}[h]
\begin{center}
\includegraphics[width=9cm]{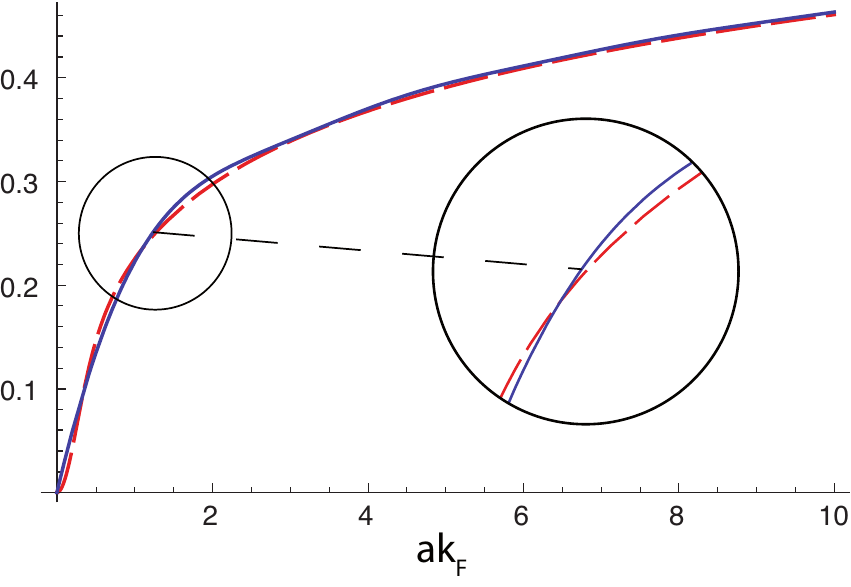}
\end{center}
\caption{\small The ground state fluctuations $\gD Q_a^2$ evaluated in the many-ferion formalism for non-interacting particles, shown here as a function of $a k_F$ (solid blue curve). The corresponding result derived by use of the boson representation is also shown (dashed red curve). The ultraviolet cutoff parameters in the two cases are identified by the relation $1/b=2.5 k_F$. \label{Fluktuasjoner2}}
\end{figure}
The corresponding expression in the boson representation is given by \pref{gsfluct2}, with the asymptotic form, for large $a/b$, given by \pref{asymp}. In the present case, with $g=1$ and with summation over the two chiralities, the asymptotic expression is 
\be{asym}
\gD Q(a,b)^2\approx \frac{1}{\pi^2}\ln(\frac{a}{b})
\ee
For the expression \pref{chfluct5} the corresponding asymptotic form for large $a k_F$ is
 \be{asym2}
\gD Q_a^2\approx \frac{1}{\pi^2}\ln(2 a k_F)
\ee
This shows that the expressions for the charge fluctuations, evaluated in these two ways, agree up to a constant if we identify the cutoff parameter $b$ in bosonized case with the inverse of the Fermi momentum in the other case. The constant can in fact be absorbed in a relative scale factor for the two cutoff parameters. This is illustrated in the figure where the charge fluctuations are plotted as functions of $\gb=a k_F$ with the identification $1/b= 2.5 \,k_F$. In spite of the fact that the high frequency contributions to the fluctuations are not identical in the two cases, the resulting curves are very close for all values of $\gb$.

\subsection{Polarization charge and fluctuations for a delta function potential}
We now turn to the polarization charge induced by an external potential which we for simplicity shall take to be a delta function. 
We examine both the expectation value and fluctuations of the local charge and compare the results to that of the low-energy bosonized description.

The Schr{\"o}dinger equation reads 
\begin{equation}
-\frac{\hbar^2}{2m} \frac{d^2}{dx^2} \psi(x) 
- \eta\delta(x-L/2)\psi(x) = E\psi(x).
\label{eq:schreq}
\end{equation}
where $\eta$ measures the strength of the delta function potential located at the point $x=L/2$, and we  take  $\eta>0$ corresponding to an attractive potential.
The delta-function potential gives rise to a discontinuity in the derivative of the wave function at $x=L/2$, and it is convenient to restrict the coordinate $x$ of the periodic variable to the interval $-L/2\leq x \leq L/2$, so that this discontinuity simply amounts to imposing the boundary conditions
\begin{align}
\dd{\psi}{x}(L/2) - \dd{\psi}{x}(-L/2)=2\eta\frac{m}{\hbar^2}\psi(L/2)\, ,
\label{eq:Deltapsidef}
\end{align}
and $\psi(-L/2)=\psi(L/2)$.

Without the potential ($\eta=0$) the energy spectrum  has a set of doubly degenerate eigenstates, which may be chosen as  symmetric (even) and antisymmetric (odd) functions,
\be{symasym}
\psi^e_n(x)=\sqrt{2\over L}\cos k_n x  \;,\quad
\psi^o_n(x)=\sqrt{2\over L}\sin k_n x   \;,\quad n=1,2,...
\ee
with wave numbers $k_n=2\pi n/ L$.  In addition there is a single non-degenerate (zero energy) state, which has $k=0$ and is therefore constant over the circle
\be{const}
\psi_0(x)={1\over\sqrt{ L}} \, .
\ee
When the interaction is turned on the zero energy state is changed to a negative energy state.  The potential thus attracts the charge of the particle in this state and forms a bound state of the form
\be{bound}
\psi_{0}(x)=A_0\cosh(\gk x) \, ,
\ee
with $\gk$ given as solution of the transcendental equation
\be{kappa}
\frac{\hbar^2}{m} \gk =\eta \coth\left(\frac{ \gk L} 2\right),
\ee
and the normalization factor given by $A_0=\sqrt{\frac{2\kappa}{\kappa L + \sinh(\kappa L)}}$.
Also the other even eigenvalue functions are modified by the potential. The form is the same as without the potential,
\be{modfunc}
\psi^e_{n}(x)=A_n\cos(\bar k_n x) \, ,
\ee
but the values of the momentum variables are shifted and now are solutions of the equation
 \be{kmod}
\frac{\hbar^2}{m} \bar k =-\eta \cot\left(\frac{ \bar k L} 2\right),
\ee
as follows from the boundary condition \pref{eq:Deltapsidef}. There is a sequence of solutions to this equation, restricted by
\be{momquant}
2\pi n/L <\bar k_n < 2\pi(n+1)/L\;,\quad n=0,1,2,...
\ee
and the modified normalization constants are
\be{norm}
A_n=\frac 1 {\sqrt L} \left[\half-\frac{\hbar^2}{m\eta L}\sin^2\left(\frac{ \bar k_n L} 2\right) \right]^{-1/2}.
\ee
The odd functions, on the other hand, are not modified by the potential since they all vanish at the point $x= L/2$. The degeneracy of the excited states are thus lifted by the potential.

We consider now the many-particle system where an odd number of particles $N=2M+1$ in the ground state occupy the $N$ lowest energy eigenstates,
\be{gstate}
\ket{G}=
\prod_{n=1}^M(c^{e\,\dag}_{n} c^{o\,\dag}_{n} )\; c^\dag_0 \ket{0} 
\ee
with $c^\dag_k$ as particle creation operators.
The field operator is expanded as
\be{fieldop2}
\psi(x)=\sum_k \psi_k(x) c_k=\sum_{n=1}^\infty \psi^e_n(x) c^e_n
+\sum_{n=1}^\infty \psi^o_n(x) c^o_n+ \psi_0(x) c_0 \, ,
\ee
and the particle number density is $\rho(x)=\psi^\dag(x)\psi(x)$. 

The ground state expectation value of the particle number density is 
\be{noexp}
\mean{\rho(x)}&=&\sum_{k\leq k_F} \psi^*_k(x)\psi_k(x)\nn
&=&\sum_{n=1}^{M} \psi^{e*}_{n}(x) \psi^{e}_{n}(x) 
+ \sum_{n=1}^{M} \psi^{o*}_{n}(x) \psi^{o}_{n}(x)
+ \psi^*_0(x) \psi_0(x) \, ,
\ee
and the density-density correlation function
\be{corr}
C(x,y)&=&\mean{(\rho(x)-\mean{\rho(x)})(\rho(y)-\mean{\rho(y)})}\nn
&=&\delta(x-y)\sum_{k\leq k_F} \psi^*_k(x)\psi_k(x)
- \sum_{k,l\leq k_F} \psi^*_k(x)\psi_l(x)\psi^*_l(y)\psi_k(y) \, .
\ee
(Note that the background charge of the unperturbed ground state has not been subtracted.) 

We use these expressions to calculate the expectation value and the variance of the local charge operator restricted to the region $D_a$ of width $a$ centered at $x= L/2$.
To find these quantities we need to evaluate numerically the following coefficients
\begin{align}
C_{nm}&= 2\int_{(L-a)/2}^{L/2} dx\,  \psi^{o*}_{n}(x) \psi^{o}_{m}(x) \nonumber \\
&= -\frac{1}{(n+m)\pi}(-1)^{n+m} \sin(\pi(n+m)a/L) \nonumber \\
&\quad +(1-\delta_{nm})\frac{1}{(n-m)\pi}(-1)^{n-m} \sin(\pi(n-m)a/L)
+ \delta_{nm}a/L
\end{align} 
\begin{align}
D_{nm}&= 2\int_{(L-a)/2}^{L/2} dx \, \psi^{e*}_{n}(x) \psi^{e}_{m}(x) \nonumber \\
&= \frac{A^*_nA_m}{{\bar k}_n+{\bar k}_m} \bigg\{ \sin(({\bar k}_n+{\bar k}_m)L/2))
-\sin(({\bar k}_n+{\bar k}_m)(L-a)/2)\bigg\} \nonumber \\
&\quad + (1-\delta_{nm}) \frac{A^*_nA_m}{{\bar k}_n-{\bar k}_m} \bigg\{ \sin(({\bar k}_n-{\bar k}_m)L/2)) 
- \sin(({\bar k}_n-{\bar k}_m)(L-a)/2)\bigg\} \nonumber \\
&\quad +\delta_{nm} A^*_nA_m \frac{a}{2} 
\end{align} 
\begin{align}
G_m&= 2\int_{(L-a)/2}^{L/2} dx \, \psi^*_{0}(x) \psi^{e}_{m}(x) \nonumber \\
&= \frac{A_0A_m}{\kappa^2 + k^2_m}
\bigg\{ {\bar k}_m ( \cosh(\kappa L/2)\sin({\bar k}_mL/2)
- \cosh(\kappa(L-a)/a)\sin({\bar k}_m(L-a)/2) ) \nonumber \\
&\qquad\qquad\qquad\quad  + \kappa ( \sinh(\kappa L/2)\cos({\bar k}_m L/2)
- \sinh(\kappa (L-a)/2)\cos({\bar k}_m (L-a)/2) ) \bigg\}
\end{align} 
\be{F}
F=2\int_{(L-a)/2}^{L/2} dx \psi^*_{0}(x) \psi_{0}(x) 
=\frac{A_0}{2\kappa} \{ \sinh(\kappa L) - \sinh(\kappa (L-a))
+\kappa a \}  \,.
\ee
The expectation value of the fermion number $N_a$ is given by
\be{mean}
\mean{N_a} = \sum_{n=1}^{M} C_{nn} + \sum_{n=1}^{M} D_{nn} + F \, ,
\ee
and the variance is
\be{var}
\Delta N^2_a = \sum_{n=1}^{M} C_{nn} + \sum_{n=1}^{M} D_{nn}  + F - F^2
- \sum_{n,m=1}^{M} C_{nm}C_{mn} + \sum_{n,m=1}^{M} D_{nm}D_{mn} 
-2\sum_{m=1}^{M} |G_m|^2 \, .\nn
\ee

In order to make a comparison with the corresponding expressions found in the bosonized formulation we include two modifications. The first one is to make a subtraction of the constant background charge of the non-interacting system and the other is to compensate for the finite value of $a/L$. The modified expression for the expectation value of the local charge is  
\be{loc}
\mean{Q_{a}}=\frac{\mean{N_a}-Na/L}{1-a/L} 
\ee
which is such that in the limit $a\rightarrow L$, $\av{Q_L} = \av{N_L} = N$.

In the numerical evaluation lengths are measured in units of $L$ and the strength of the potential in is measured by the dimensionless parameter $\tilde\eta=(mL/\hbar^2)\eta$. The wave numbers $\bar k_n$ have been determined by solving numerically Eq.\pref{kmod} and the value of $\gk$ has similarly been found by numerically solving \pref{kappa}. These values have been used when evaluating the coefficients $D_{nm}$, $G_m$ and $F$. 

\begin{figure}[h]
\begin{center}
\includegraphics[width=12cm]{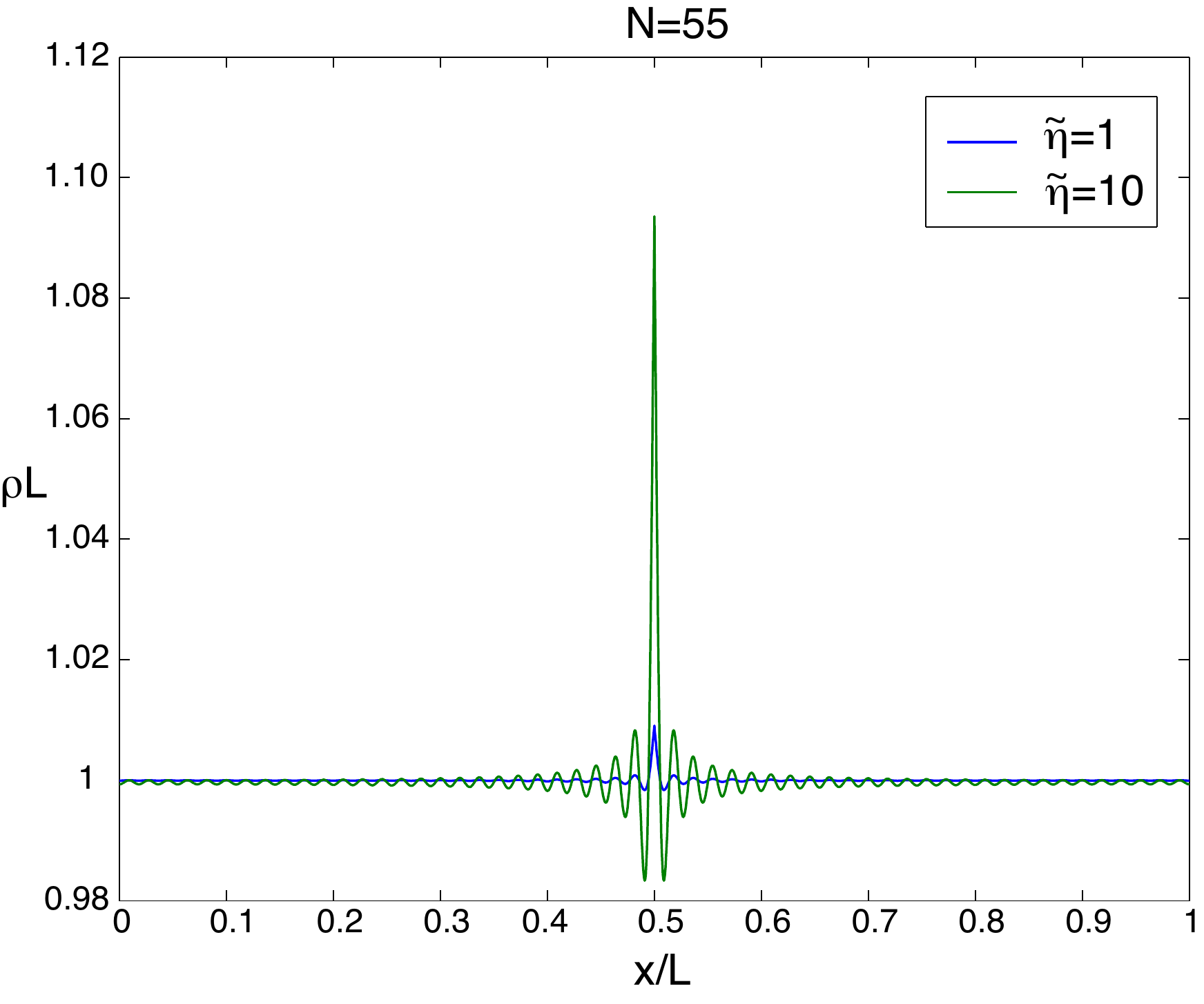}
\end{center}
\caption{\small Charge density in the background of a delta function potential. The charge distribution is plotted for $N=55$ particles and for two values $\tilde\eta=1$ and $\tilde\eta=10$ of the potential strength parameter.
The density shows in both cases a peak at the location of the potential, with largest amplitude for the largest value of $\tilde\eta$. The oscillations are a high frequency effect, with wave length determined by the Fermi momentum $k_F$. \label{dens55}}
\end{figure}

The numerically evaluated ground state expectation value of the charge density is shown in Fig.~6 for particle number $N=55$ and for two values of the potential strength $\tilde\eta$. The polarization charge induced by the delta function potential is strongly localized around the point $x= L/2$, and more so for larger than for the smaller value of $\tilde\eta$. The high frequency Friedel oscillations in the density occur at the wave vector    $2k_F=2\pi N/L$ and their amplitude decrease with increasing $N$.   $k_F $ also determines the width of the central peak, and this is consistent with the Fermi momentum effectively defining an ultraviolet cutoff.

\begin{figure}[h]
\begin{center}
\includegraphics[width=10cm]{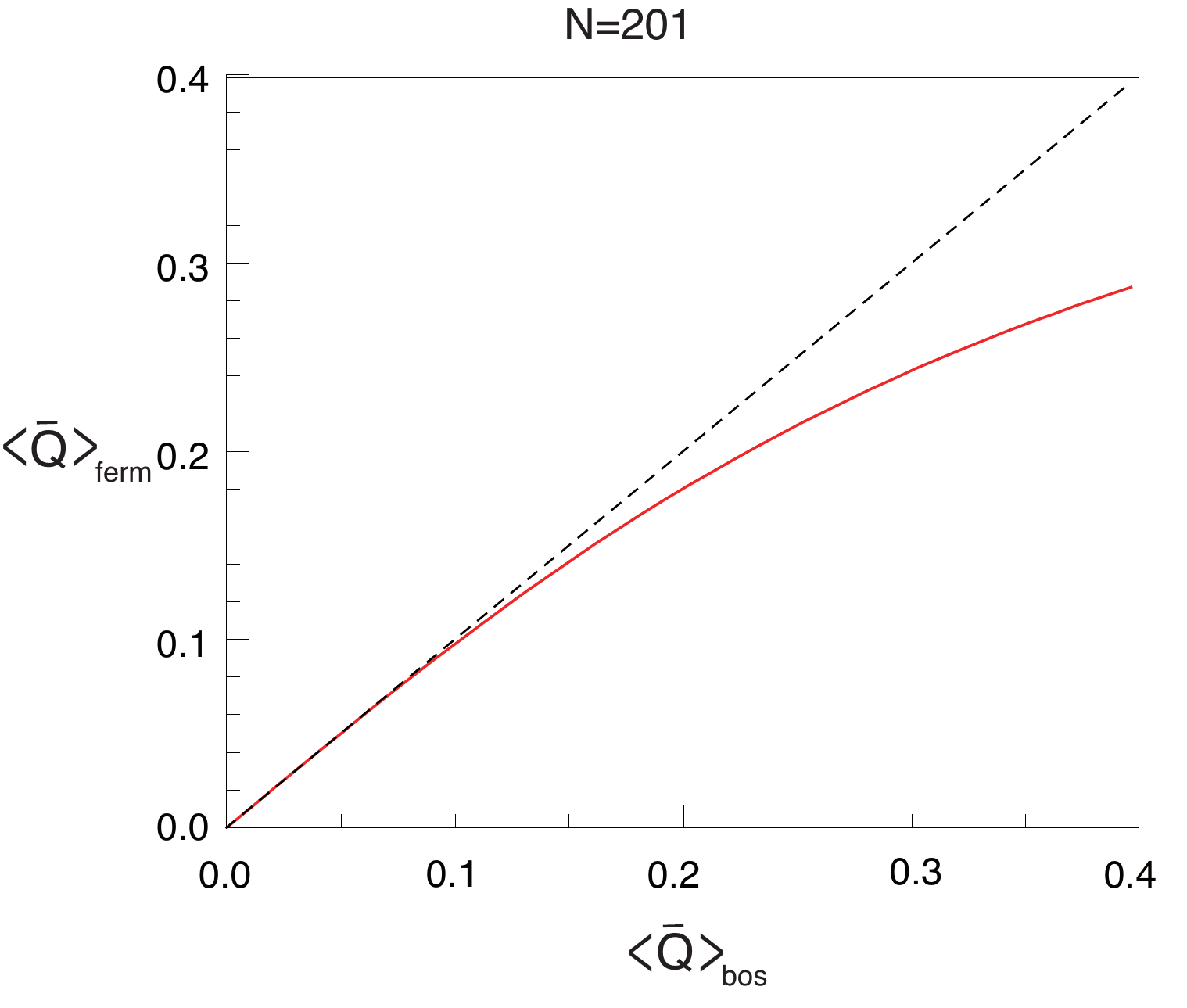}
\end{center}
\caption{\small Expectation of the total local charge, $\mean{\bar Q}_{ferm}$, determined by the many-fermion calculation, shown as a function of the charge $\mean{\bar Q}_{bos}$ determined by calculation in the boson representation. The result is shown by the solid (red) curve for $N=201$ particles. The dotted curve shows the line of equality for the two charges. For small charge values (weak potential) the two methods to calculate the charge agree well, while for larger charge values (strong potential) they give diverging results. The difference can be ascribed to the effect of having a finite value for $k_F$ in the fermion calculation.  \label{Ladning}}
\end{figure}

To compare the results with those found in the boson representation we focus on the value of the total local charge $\bar Q=\sum_\gG \bar Q_\gG$. In the non-interacting case, with $g=1$ and $W_0=-\eta$ for the strength of the potential, the value previously found for the charge is
\be{tot}
\mean{\bar Q}_{bos}  = \frac{\eta}{\pi \hbar u}=\frac{1}{\pi^2}\frac{\tilde\eta}{N}
\ee
We have labeled the charge with "bos" to specify that it is evaluated in the low energy, bosonized approximation. The corresponding charge evaluated by using the many-fermion formalism is labeled $\mean{\bar Q}_{ferm}$, where this charge is identified with $\mean {Q_a}$ for a value of $a$ that is sufficiently large to capture fully the contribution from the central part of the charge distribution. 

In Fig.~7 the charge $\mean{\bar Q}_{ferm}$ is shown as function of $\mean{\bar Q}_{bos}$. The curve is obtained by varying the potential strength $\tilde\eta$ with fixed particle number, in the figure corresponding to $N=201$. 
The curve shows that $\mean{\bar Q}_{ferm}$ and $\mean{\bar Q}_{bos}$ agree well for small values $\mean{\bar Q}_{ferm}\lesssim 0.1$, corresponding to a weak potential strength $\eta$ (measured relative to the Fermi velocity). When the strength of the potential increases, however there is an increasing discrepancy. This has to be viewed as effects of high frequency contributions, which are treated differently by the two methods. To some extent this can be seen as a consequence of using a delta function potential, since this has Fourier components of arbitrary high frequency. 

The increase in importance of these high frequency components with the strength of the potential can be understood in the boson representation in the following way. The quantum state  in the background of the potential is a coherent state which can be expanded in powers of the operator $(\frac{\Delta_q}{\omega_q}b_q^\dag)^n$ that act on the ground state $\ket G$. Since ${\Delta_q}$ is proportional to the potential strength $\eta$ this shows that higher powers of $b_q^\dag$, and therefore higher energy contributions, will be more important for large than for small values of $\eta$.

\begin{figure}[h]
\begin{center}
\includegraphics[width=12cm]{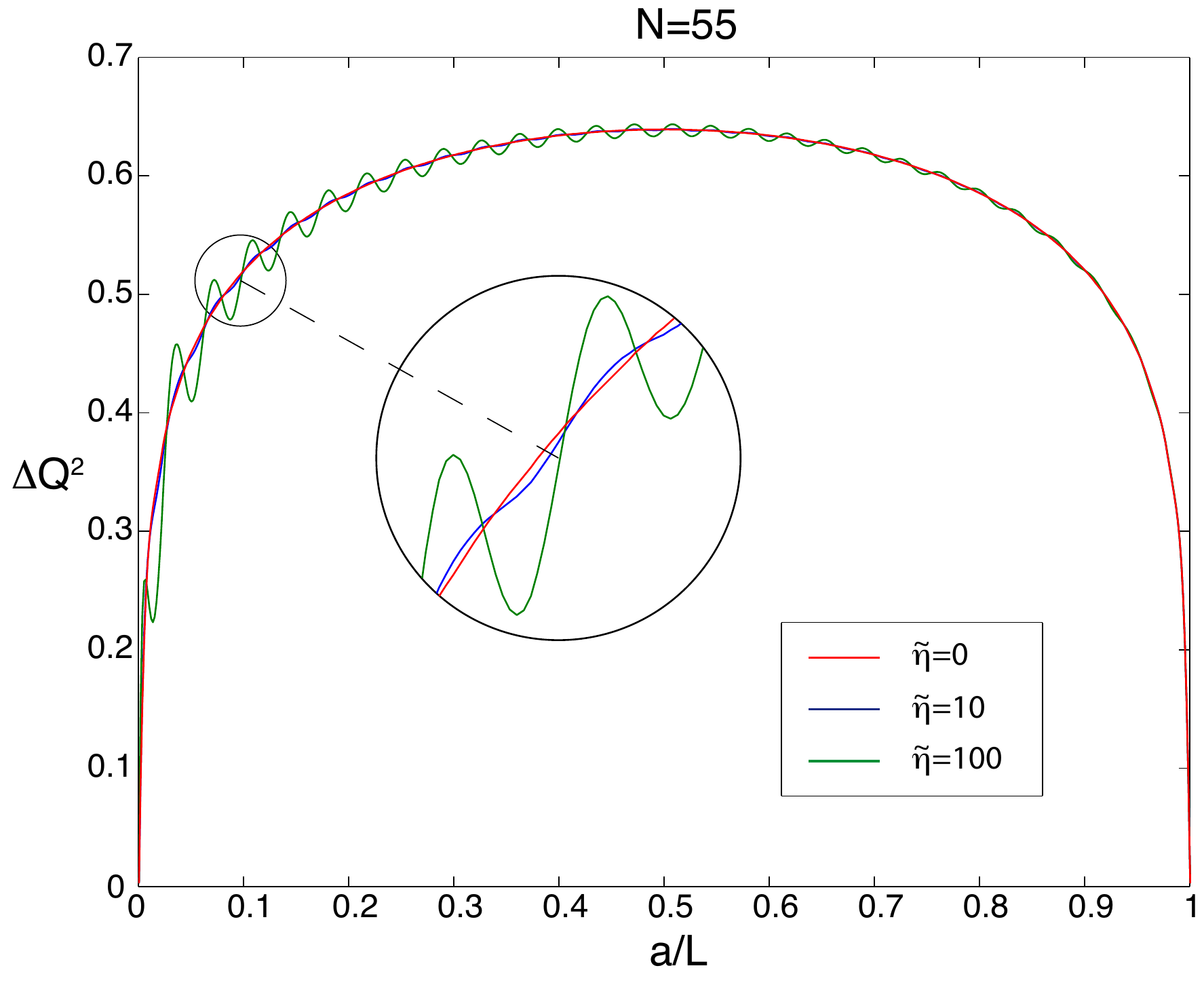}
\end{center}
\caption{\small Charge fluctuations in the background potential. The plot shows the variance $\gD Q_a^2$ as a function of $a/L$ for three different values of the potential strength $\tilde\eta$. The particle number is in all three  cases $N=55$. The deviations from the ground state fluctuations are very small for $\tilde\eta=10$, but can be seen in the window with magnification of a section of the curves.  \label{fluct55}}
\end{figure}

The charge fluctuations $\gD Q_a^2$ have been evaluated for different values of $a/L$, $\tilde\eta$ and $N$. The interesting point has been to check if, also when evaluated by use of the many-fermion wave function,  the fluctuations are identical to those of the ground state, so that this result is not just an artifact of the low energy approximation. In Fig.~8 the results are plotted as functions of $a/L$ for $N=55$ and for three different values of $\tilde\eta$, one of them corresponding to the value $\tilde\eta=0$ of the free system. As shown by the plot there are deviations from this curve for non-vanishing values of $\tilde\eta$, but these have the character of oscillations about the fluctuations curve of the free system. The oscillations seem to be rapidly damped when $\tilde\eta$ decreases, and in the plot that is demonstrated by the difference in the amplitude of the oscillations for the curves with $\tilde\eta=10$ and $\tilde\eta=100$. 

It is of interest to compare the results displayed in Fig.~8 with those in Fig.~7, where the results for the evaluations of the expectation value of the charge are shown.  The discrepancy between the results for $\mean{\bar Q}_{ferm}$ and $\mean{\bar Q}_{bos}$ shown  in Fig.~7 can be ascribed to the effect of the finite value of the Fermi momentum $k_F$ for the fermion result, and the curves indicate that this effect decreases rapidly for charge values below $\mean{\bar Q}\approx 0.1$. For the particle number $N=55$, as used  in  Fig.~8, the parameter value $\tilde\eta=10$  corresponds to $\mean{\bar Q}\approx 0.02$ and  $\tilde\eta=100$ corresponds to $\mean{\bar Q}\approx 0.2$. The curves shown in this figure demonstrate that also the oscillations of the charge fluctuations are rapidly damped below $\mean{\bar Q}\approx 0.1$. It seems natural to assume that the appearance of these oscillations are also due to the finite value of $k_F$, and a closer inspection of the periodicities confirms this assumption.
The oscillations in the fluctuations are then closely related to the oscillations in the charge density displayed in Fig.~6, and they therefore appear as a consequence of the use a sampling function with sharp edges, since a smooth edge would suppress high frequency contributions with $k\approx k_F$. Apart from these oscillations we find no significant difference between the fluctuations of the local charge  and the fluctuations of the ground state.

Let us finally stress the point that the problems met here in getting a simpler picture of the charge fluctuations are linked to our use of a delta-function potential and a sampling function with sharp edges. Both these functions could be smoothened, but that would involve a more demanding calculation than the one we have aimed at in this Appendix.  
 
\end{document}